\documentclass[final,3p,times,number]{elsarticle}
\usepackage{xcolor}
\usepackage{amsmath}
\usepackage{natbib}
\biboptions{numbers,square,sort&compress}
\usepackage{mathtools}
\usepackage{physics}
\usepackage{stmaryrd} 
\usepackage{epstopdf}
\usepackage{grffile}   
\usepackage{bm}
\usepackage{graphicx,wrapfig}
\usepackage{wrapfig}
\usepackage[T1]{fontenc}
\usepackage[normalem]{ulem}
\usepackage{algorithm}
\usepackage{algpseudocode}
\usepackage{tikz}
\usetikzlibrary{shapes.geometric, arrows.meta, positioning, fit, backgrounds}

\newcommand{\stkout}[1]{\ifmmode\text{\sout{\ensuremath{#1}}}\else\sout{#1}\fi}
\usepackage{mathrsfs}
\usepackage{amssymb}
\usepackage{nomencl}
\usepackage{etoolbox}
\usepackage{multicol}
\usepackage{caption,subcaption}
\usepackage{tabularx}
\usepackage{booktabs}
\usepackage{multirow}
\makenomenclature

\usepackage{cleveref}
\crefname{equation}{Eq.}{Eqs.}
\Crefname{equation}{Eq.}{Eqs.}

\begin{document}

\begin{frontmatter}


\title{PDE-constrained optimization for virtual sensing in structural dynamics: Full-field displacement and force recovery from sparse sensors}

\author[1]{Minjae Kim\fnref{equal}}
\author[1]{Jaehwan Jeong\fnref{equal}}
\fntext[equal]{These authors contributed equally to this work.}


\author[2]{Seulki Lee}

\author[1]{Jaemin Kim\corref{cor1}}
\cortext[cor1]{Corresponding Author}
\ead{jaeminkim@changwon.ac.kr}

\address[1]{School of Mechanical Engineering, Changwon National University, Changwon 51140, Republic of Korea \fnref{label1}}
\address[2]{Department of Architectural Engineering, Kwangwoon University, Seoul 01897, Republic of Korea}

\begin{abstract}
Virtual sensing --- recovering full-field structural response from sparse sensor measurements --- is a fundamental challenge in structural health monitoring (SHM).
Fracture assessment, fatigue evaluation and remaining useful life (RUL) prognosis all take the applied load as their input, and that load frequently acts on surfaces that cannot be instrumented. An improved load estimate is therefore of direct practical value for SHM and for prognostics and health management (PHM).
The mode-based approach reconstructs the displacement field accurately from a small number of sensors and has been applied successfully for that purpose.
The load inferred from this reconstructed displacement is far less reliable, because the residual displacement error is amplified when it is mapped back through the dynamic stiffness.
This study therefore formulates virtual sensing as a PDE-constrained optimization (PDE-CO) problem in which the elastodynamic equation is an equality constraint and the applied load is the optimization variable, so that displacement and load are estimated together rather than in sequence.
To manage the computational cost, the framework separates offline finite element model preparation from online reconstruction.
The assembled matrices are reused during online optimization, and the forward and gradient evaluations can use GPU-based parallel computation.
The numerical comparisons support improved noise robustness of PDE-CO displacement reconstruction relative to the mode-based approach within the tested settings.
These results provide a theoretical and computational foundation for the development of virtual sensing for SHM and PHM.
\end{abstract}

\end{frontmatter}

\section*{Nomenclature}
\begingroup
\linespread{1}\footnotesize
\setlength{\tabcolsep}{2.8pt}
\renewcommand{\arraystretch}{1.05}
\begin{center}
\begin{tabular*}{\textwidth}{@{\extracolsep{\fill}}llll@{}}
\hline
Symbol & Definition & Symbol & Definition \\
\hline
$\mathbf{K},\mathbf{M}$ & Stiffness and consistent mass matrices & $\mathbf{C}_{d,ff}$ & Damping matrix for free DOFs \\
$\mathbf{D}(\omega),\mathbf{D}_{ff}$ & Dynamic stiffness and its free block & $\mathbf{P},\mathbf{S}_f$ & Force mapping and sensor selection \\
$\boldsymbol{\Phi},\mathbf{q}$ & Retained modes and modal amplitudes & $\boldsymbol{\theta}$ & Unknown force parameters \\
$\hat{\mathbf{U}}_f$ & Displacement amplitudes at free DOFs & $\hat{\mathbf{a}}_{\mathrm{obs}},\hat{\mathbf{a}}_{\mathrm{pred}}$ & Measured and predicted acceleration amplitudes \\
$\boldsymbol{\varepsilon},\boldsymbol{\sigma}$ & Small strain and Cauchy stress tensors & $J(\boldsymbol{\theta}),\alpha$ & Objective and regularisation weight \\
$E,\nu$ & Young's modulus and Poisson's ratio & $\rho,\mathbb{C}$ & Mass density and elasticity tensor \\
$\omega,\omega_i$ & Excitation and modal frequencies (rad/s) & $\Omega$ & Physical domain \\
$N,N_f,N_s$ & Total, free and sensor DOF counts & $N_\theta,N_d$ & Counts of force parameters and modes \\
$\Gamma_D,\Gamma_N$ & Dirichlet and Neumann boundaries & $\mathbf{A}$ & Maps force to sensor acceleration \\
$\mathcal{M}$ & Finite set of structural models & $\mathcal{S}$ & Criterion for sensor placement \\
\hline
\end{tabular*}
\end{center}
\endgroup

\section{Introduction}\label{sec:1}

\begin{figure}[!ht]
\centering
\includegraphics[width=0.85\textwidth]{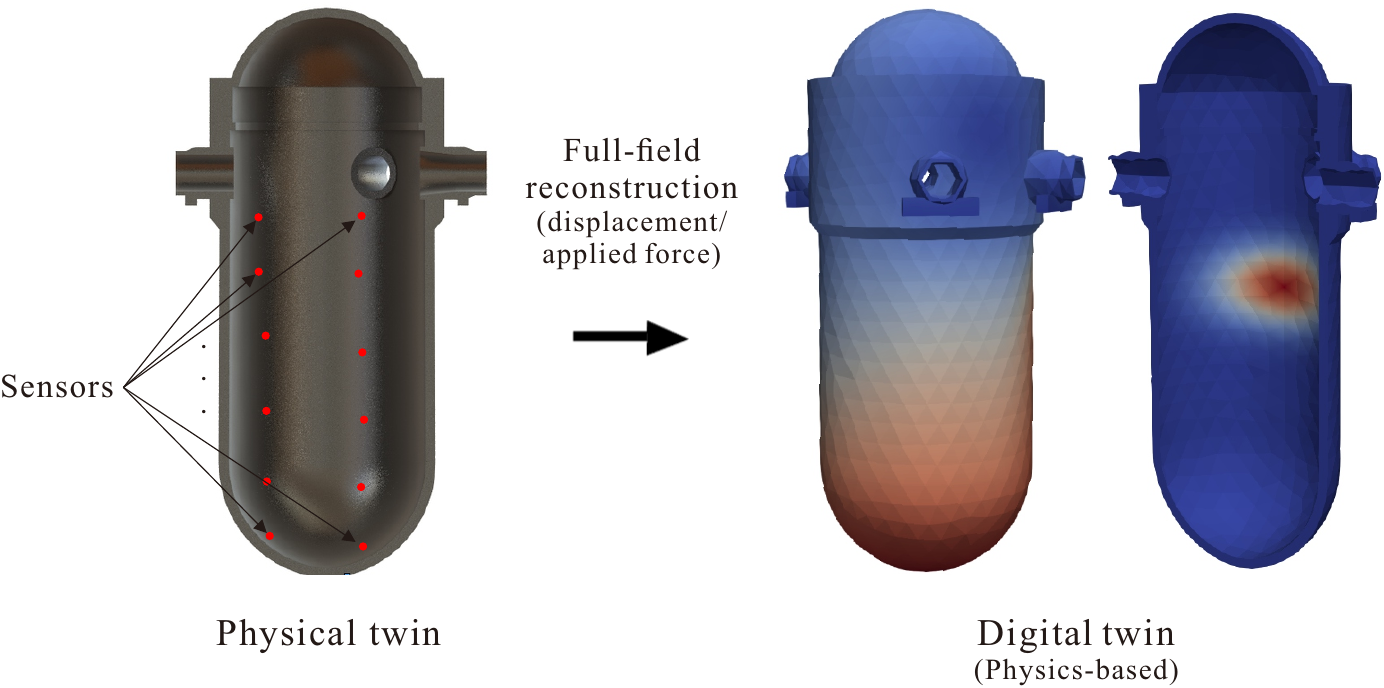}
\caption{Conceptual virtual-sensing arrangement. Sparse sensors on the physical structure are linked to a physics-based numerical model for displacement and applied-force inference. The diagram illustrates the intended information flow; the numerical examples in this paper do not constitute experimental validation of a physical twin.}\label{fig:schematic}
\end{figure}

Virtual sensing reconstructs structural responses at unmeasured locations from sparse sensor measurements (Figure~\ref{fig:schematic}) \citep{lee2024virtual,mora2023strain}.
It supports structural health monitoring (SHM) and fatigue assessment \citep{maes2016dynamic,tarpo2020expansion}.
More broadly, combining structural models with measurements enables damage detection using digital twins, model updating and modal properties \citep{ritto2021digital,lee2023structural,ge2005structural}.
Alongside the response field, the applied load is an important input to fracture assessment, fatigue evaluation and remaining useful life (RUL) prediction.
In practice, loads may act on internal or inaccessible surfaces where direct instrumentation is difficult.
Estimating these loads from accessible response measurements can therefore support both SHM and prognostics and health management (PHM).
For this purpose, the quality of virtual sensing must be assessed in terms of the recovered load as well as the reconstructed displacement.

Virtual-sensing methods include modal least-squares estimation and Kalman filtering \citep{lee2024virtual,mora2023strain,lourens2012augmented}.
Related reduced-order modelling developments improve the accuracy of Craig--Bampton component mode synthesis by accounting for residual modes \citep{kim2017improving,kim2018considering}.
Recent work also combines physics-guided learning with Kalman filtering for virtual sensing of nonlinear structures \citep{haywood2026piggo}.
The mode-based approach considered here approximates displacement as a linear combination of a limited number of low-order modes and estimates their coordinates from sensor data \citep{fu2001modal,iliopoulos2016modal}.
This approach can reconstruct displacement accurately from a small number of sensors.
The downside of this approach is that accurate displacement reconstruction does not necessarily provide an accurate load estimate.
For the mode-based approach considered here, the load is computed by applying the dynamic stiffness matrix to the reconstructed displacement.
Although the resulting load satisfies discrete equilibrium, this operation can amplify displacement errors caused by modal truncation and measurement noise.
The regularization used in this estimator controls the modal coordinates but does not directly restrict the load to a specified region or penalize its magnitude.
For these reasons, this study adopts an inverse formulation that estimates the applied load directly, with explicit constraints on its location and a penalty on its magnitude.

In this article, the analysis is limited to linear harmonic excitation, and virtual sensing is formulated as a PDE-constrained optimization (PDE-CO) problem.
The applied load is treated as the unknown, and the elastodynamic equation is imposed as an equality constraint that relates the load to the displacement field.
At each optimization step, the finite element model predicts the response to a candidate load, and the load is updated to reduce the discrepancy between predicted and measured sensor responses.
The load is restricted to a specified candidate region, and regularization penalizes large load amplitudes.
This formulation estimates the load and its corresponding displacement within the same inverse problem by solving the full finite element equations directly, without relying on a reduced modal basis.
Sparse measurements may still be consistent with different load distributions, so the reconstruction depends on the assumed load region and regularization.

PDE-CO has been applied to a range of inverse and design problems in computational mechanics, including elasticity imaging, structural weakness identification, constitutive inverse design and structural sizing under dynamic loads \citep{oberai2003inverse,kostic2023adjoint,bahmani2026constitutive,marsden2021adjoint}.
These applications determine material parameters, defects or design variables by minimizing a measurement discrepancy or design objective subject to the governing field equations.
Adjoint methods and differentiable simulation provide the sensitivities required for gradient-based optimization \citep{xue2023jaxfem,bezgin2023jaxfluids}.
PDE-CO has also been used to identify excitation sources in elastodynamics and acoustics for reproducing acoustic and vibration test conditions \citep{walsh2013source}.

To the best of the authors' knowledge, this is the first application of a full-order PDE-CO formulation with automatic differentiation to structural virtual sensing for joint load and displacement recovery from sparse acceleration measurements under harmonic excitation.
The key features of the proposed framework include (i) full-order finite element equilibrium for joint load and displacement reconstruction, (ii) a prescribed candidate load region and regularization of load amplitudes, and (iii) offline model assembly with online optimization supported by automatic differentiation and GPU-based computation.

Three synthetic numerical examples, comprising a cantilever plate, a $90^\circ$ elbow pipe and a reactor pressure vessel, compare displacement reconstruction with the mode-based approach. In matched-model tests, adding 3\% acceleration noise increases the displacement error at non-sensor nodes by 1.18--2.15 percentage points for PDE-CO, compared with 3.90--79.86 for the mode-based approach with the original retained number of modes.

These results establish the practical significance and limitations of the formulation for load and response estimation in SHM and PHM.
The remainder of this paper is organized as follows.
Section~\ref{sec:2} reviews the mode-based approach and its limitations, and Section~\ref{sec:3} introduces PDE-constrained optimization as an alternative inverse framework.
Section~\ref{sec:4} develops the harmonic PDE-CO formulation, including the Tikhonov-regularized objective, the force parameterization, the gradient computation and the output-specific reliability assessment.
Section~\ref{sec:6} describes the implementation, which couples an offline FEniCSx assembly stage with an online JAX optimization stage and a sparse direct reference solution.
Section~\ref{sec:7} reports the numerical experiments and additional validation studies, and Section~\ref{sec:8} summarizes the findings and limitations.

Throughout this paper, bold symbols denote vectors, matrices or tensors; $\top$ denotes transpose, and $\|\cdot\|_2$ is the Euclidean norm for vectors. Subscripts $f$ and $c$ identify free and constrained degrees of freedom. A hat denotes a harmonic amplitude, while superscripts $\mathrm{est}$ and $*$ identify an estimated quantity and an optimization solution, respectively. Modal coordinates $\mathbf q$ and load parameters $\boldsymbol\theta$ are defined directly in the amplitude domain and omit the hat. The objective is denoted by $\mathcal J$, and $\widetilde{\mathcal{J}}$ is its reduced form after eliminating the displacement. The scalar regularization penalty $\mathcal R$ is distinguished from the vector governing-equation residual $\mathbf R$. Other symbols are defined when introduced and summarized in the nomenclature.

\section{Mode-based approach for virtual sensing and its limitations}\label{sec:2}

\subsection{Mode-based reconstruction procedure}\label{sec:2.1}

The mode-based approach reconstructs the full displacement field from sparse sensor measurements \citep{fu2001modal,iliopoulos2016modal}.
The method begins by solving the generalized eigenvalue problem associated with the free-DOF system:
\begin{align}\label{eq:gep}
    \mathbf{K}_{ff}\,\boldsymbol{\phi}_i = \omega_i^2\,\mathbf{M}_{ff}\,\boldsymbol{\phi}_i, \quad i = 1, \ldots, N_f,
\end{align}
where $\omega_i$ is the $i$-th natural frequency and $\boldsymbol{\phi}_i \in \mathbb{R}^{N_f}$ is the corresponding eigenmode.
The eigenmodes are normalized to be mass-orthonormal, so the modal matrix $\boldsymbol{\Phi} = [\boldsymbol{\phi}_1, \ldots, \boldsymbol{\phi}_{N_d}] \in \mathbb{R}^{N_f \times N_d}$ composed of the first $N_d$ dominant modes satisfies the orthogonality conditions:
\begin{align}\label{eq:modal_orthogonality}
    \boldsymbol{\Phi}^\top \mathbf{M}_{ff}\, \boldsymbol{\Phi} = \mathbf{I}_{N_d}, \quad
    \boldsymbol{\Phi}^\top \mathbf{K}_{ff}\, \boldsymbol{\Phi} = \boldsymbol{\Lambda},
\end{align}
where $\boldsymbol{\Lambda} = \mathrm{diag}(\omega_1^2, \ldots, \omega_{N_d}^2)$ is the spectral matrix.

The displacement amplitude of the free DOFs is approximated using the first $N_d$ modes:
\begin{align}\label{eq:modal_superposition}
    \hat{\mathbf{U}}_f \approx \boldsymbol{\Phi}\,\mathbf{q},
\end{align}
where $\mathbf{q} \in \mathbb{R}^{N_d}$ is the modal coordinate vector to be determined from sensor data.

Under harmonic excitation at frequency $\omega$, the sensor acceleration data $\hat{\mathbf{a}}_{\mathrm{obs}} \in \mathbb{R}^{N_s}$ is first converted to sensor displacements via the kinematic relationship $\hat{\mathbf{a}} = -\omega^2 \hat{\mathbf{u}}$:
\begin{align}\label{eq:modal_u_from_a}
    \hat{\mathbf{u}}_{\mathrm{obs}} = -\frac{\hat{\mathbf{a}}_{\mathrm{obs}}}{\omega^2}.
\end{align}
The sensor displacements are related to the full free-DOF displacement through the sensor selection matrix $\mathbf{S}_f \in \mathbb{R}^{N_s \times N_f}$ (a Boolean matrix that extracts sensor DOFs):
\begin{align}\label{eq:modal_sensor_extraction}
    \hat{\mathbf{u}}_{\mathrm{obs}} = \mathbf{S}_f\,\hat{\mathbf{U}}_f \approx \mathbf{S}_f\,\boldsymbol{\Phi}\,\mathbf{q} = \boldsymbol{\Phi}_m\,\mathbf{q},
\end{align}
where $\boldsymbol{\Phi}_m \coloneqq \mathbf{S}_f\,\boldsymbol{\Phi} \in \mathbb{R}^{N_s \times N_d}$ is the modal matrix restricted to sensor DOFs.

\subsection{Full-field reconstruction}\label{sec:2.2}

The modal coordinates $\mathbf{q}$ are recovered by minimizing the Tikhonov-regularized least-squares cost function:
\begin{align}\label{eq:modal_cost}
    \mathcal{J}_{\mathrm{modal}}(\mathbf{q}) = \| \boldsymbol{\Phi}_m\,\mathbf{q} - \hat{\mathbf{u}}_{\mathrm{obs}} \|^2 + \alpha_m \| \mathbf{q} \|^2,
\end{align}
where $\alpha_m > 0$ is a regularization parameter that penalizes large modal coordinates.
Setting $\nabla_{\mathbf{q}} \mathcal{J}_{\mathrm{modal}} = \mathbf{0}$ yields the modal-coordinate estimate $\mathbf q^{\mathrm{est}}$:
\begin{align}\label{eq:modal_recovery}
    \mathbf{q}^{\mathrm{est}} = (\boldsymbol{\Phi}_m^\top\boldsymbol{\Phi}_m + \alpha_m\,\mathbf{I}_{N_d})^{-1}\boldsymbol{\Phi}_m^\top\,\hat{\mathbf{u}}_{\mathrm{obs}}.
\end{align}
When $N_s \geq N_d$ and $\boldsymbol{\Phi}_m$ has full column rank and acceptable conditioning, the unregularized pseudoinverse ($\alpha_m = 0$) can be used. Sensor count alone does not establish this rank condition, and positive regularization also permits $N_d>N_s$.
In practice, however, the number of available sensors is often limited ($N_s \leq N_d$) or the system is ill-conditioned, making regularization essential for a stable solution.

The full-field displacement at all free DOFs is then reconstructed as:
\begin{align}\label{eq:modal_full}
    \hat{\mathbf{U}}_f^{\mathrm{modal}} = \boldsymbol{\Phi}\,\mathbf{q}^{\mathrm{est}}.
\end{align}
The inversion involves only an $N_d \times N_d$ system, making the mode-based approach computationally efficient and widely used for real-time virtual sensing.
The retained mode shapes are derived from the governing structural model, so the mode-based approach is physics-based. The least-squares fit in \cref{eq:modal_cost} uses the measured response alone and imposes neither a prescribed force support nor a force-space prior. The mode-based approach is in this sense a response-only estimator, as opposed to the input-based formulations of Section~\ref{sec:3}.
The mode-based approach recovers an approximate displacement field but with accuracy limited by modal truncation error.

\subsection{Limitations: force recovery from modal displacement}\label{sec:2.3}

A natural question therefore arises: can the applied force be recovered accurately once the displacement field has been reconstructed?
Given the free-DOF dynamic stiffness matrix $\mathbf{D}_{ff} = \mathbf{K}_{ff} - \omega^2\mathbf{M}_{ff}$ (derived in Section~\ref{sec:4.discretization}), the force estimate from the mode-based approach is:
\begin{align}\label{eq:modal_force}
    \hat{\mathbf{F}}_f^{\mathrm{modal}} = \mathbf{D}_{ff}\, \hat{\mathbf{U}}_f^{\mathrm{modal}}.
\end{align}
This can produce an inaccurate force estimate for two reasons, even when the displacement error is small.

The first is error amplification through the dynamic stiffness operator.
The true displacement can be expanded in the complete set of $N_f$ eigenmodes as $\hat{\mathbf{U}}_f^{\mathrm{true}} = \sum_{i=1}^{N_f} \boldsymbol{\phi}_i\, q_i^{\mathrm{true}}$, where $q_i^{\mathrm{true}}$ is the true modal coordinate for mode $i$.
The mode-based approach retains only the first $N_d$ modes, so $\hat{\mathbf{U}}_f^{\mathrm{modal}} = \hat{\mathbf{U}}_f^{\mathrm{true}} + \boldsymbol{\epsilon}$, where the truncation error is:
\begin{align}\label{eq:truncation_error}
    \boldsymbol{\epsilon} = \sum_{i=1}^{N_d}(q_i^{\mathrm{est}} - q_i^{\mathrm{true}})\,\boldsymbol{\phi}_i - \sum_{i=N_d+1}^{N_f} \boldsymbol{\phi}_i\, q_i^{\mathrm{true}}.
\end{align}
The first sum represents the least-squares fitting error (typically small), and the second sum represents the modal truncation error from omitted higher modes.
Applying the dynamic stiffness operator yields the force error $\hat{\mathbf{F}}_f^{\mathrm{modal}} = \hat{\mathbf{F}}_f^{\mathrm{true}} + \mathbf{D}_{ff}\,\boldsymbol{\epsilon}$.
Since the eigenmodes satisfy $\mathbf{K}_{ff}\,\boldsymbol{\phi}_i = \omega_i^2\,\mathbf{M}_{ff}\,\boldsymbol{\phi}_i$ and $\mathbf{D}_{ff}\,\boldsymbol{\phi}_i = (\omega_i^2 - \omega^2)\,\mathbf{M}_{ff}\,\boldsymbol{\phi}_i$, the force error becomes:
\begin{align}\label{eq:force_amplification_detail}
    \mathbf{D}_{ff}\,\boldsymbol{\epsilon}
    = \underbrace{\sum_{i=1}^{N_d}(q_i^{\mathrm{est}} - q_i^{\mathrm{true}})\,(\omega_i^2 - \omega^2)\,\mathbf{M}_{ff}\boldsymbol{\phi}_i}_{\text{amplified fitting error}}
    \;-\;\underbrace{\sum_{i=N_d+1}^{N_f} (\omega_i^2 - \omega^2)\,q_i^{\mathrm{true}}\,\mathbf{M}_{ff}\boldsymbol{\phi}_i}_{\text{truncated-mode contribution}}.
\end{align}
The magnitude of $(\omega_i^2 - \omega^2)$ increases for omitted modes sufficiently above the excitation frequency. Consequently, a small omitted displacement contribution $\boldsymbol{\phi}_i q_i^{\mathrm{true}}$ may correspond to an appreciable force contribution in \cref{eq:force_amplification_detail}. The size of this effect depends on the load spectrum, mass normalization and distance from resonance; it does not imply a universal decay or frequency-scaling law.
Higher modes may therefore contribute negligibly to the displacement field but significantly to the force field.
This should not come as a surprise since the dynamic stiffness operator can amplify small displacement contributions from omitted high-order modes.

The second reason is that a local force is being represented in a global basis.
The true load is applied on $\Gamma_N$ and has at most $N_\theta$ nonzero entries, whereas every $\mathbf{M}_{ff}\boldsymbol{\phi}_i$ in \cref{eq:modal_force} is nonzero over the whole domain and \cref{eq:modal_cost} imposes no force-related constraint. Any combination of the retained modes is therefore generically nonzero at every free degree of freedom, so the estimate assigns load outside the loaded region however well the displacement is fitted. Together with the amplification in \cref{eq:force_amplification_detail}, this makes the load estimate from the mode-based approach sensitive to measurement noise (Section~\ref{sec:7}).

\section{PDE-constrained optimization as an alternative}\label{sec:3}

The limitations identified in Section~\ref{sec:2} --- modal truncation error and the absence of an explicit force-support prior in the mode-based approach --- motivate an alternative parameterization. PDE-constrained optimization (PDE-CO) enforces the governing equation as an equality constraint, operates on the full-order FE system without modal reduction, and estimates the unknown force distribution by minimizing a regularized sensor misfit. These choices remove basis truncation from the forward solve but do not remove the inverse problem's null space or the FE model error.

\subsection{Key idea and general formulation}\label{sec:3.1}

The advantages of PDE-CO over the mode-based approach are summarized in the Introduction. The key idea is to treat the unknown force $\hat{\mathbf{F}}_f \in \mathbb{R}^{N_f}$ as the optimization variable and enforce the governing PDE as an equality constraint.
Given a trial force $\hat{\mathbf{F}}_f$, the forward solve yields the displacement:
\begin{align}\label{eq:pdeco_forward}
    \hat{\mathbf{U}}_f = \mathbf{D}_{ff}^{-1}\,\hat{\mathbf{F}}_f,
\end{align}
where $\mathbf{D}_{ff} = \mathbf{K}_{ff} - \omega^2\mathbf{M}_{ff}$ is the dynamic stiffness matrix (derived in Section~\ref{sec:4.discretization}).
The predicted response at the sensor locations is then extracted as $\hat{\mathbf{a}}_{\mathrm{pred}} = -\omega^2\,\mathbf{S}_f\,\hat{\mathbf{U}}_f$, where $\mathbf{S}_f \in \mathbb{R}^{N_s \times N_f}$ is a Boolean selection matrix.
The optimization minimizes the misfit between the predicted and observed sensor accelerations, $\hat{\mathbf{a}}_{\mathrm{pred}}$ and $\hat{\mathbf{a}}_{\mathrm{obs}}$:
\begin{align}\label{eq:pdeco_objective_overview}
    \min_{\hat{\mathbf{F}}_f} \; \mathcal{J} = \bigl\|\hat{\mathbf{a}}_{\mathrm{pred}} - \hat{\mathbf{a}}_{\mathrm{obs}}\bigr\|^2 + \alpha\,\bigl\|\hat{\mathbf{F}}_f\bigr\|^2,
\end{align}
where $\alpha$ is the Tikhonov regularization weight.
The PDE constraint in \cref{eq:pdeco_forward} is embedded in the objective through the forward solve, so any minimizer automatically satisfies the governing equation.
More generally, the force is parameterized as $\hat{\mathbf{F}}_f = \mathbf{P}\boldsymbol{\theta}$, where $\mathbf{P} \in \mathbb{R}^{N_f \times N_\theta}$ is a parameterization matrix that selects or maps the optimization variable $\boldsymbol{\theta} \in \mathbb{R}^{N_\theta}$ to the nodal force vector.
When the loaded boundary $\Gamma_N$ is known, $\mathbf{P}$ is a Boolean selection matrix that restricts forces to the DOFs on $\Gamma_N$, with $N_\theta$ equal to the number of candidate loaded DOFs.

Substituting $\hat{\mathbf F}_f=\mathbf P\boldsymbol\theta$ into the load penalty in \cref{eq:pdeco_objective_overview} defines its expression in the force parameters:
\begin{equation}\label{eq:force_penalty_mapping}
 \mathcal R(\boldsymbol\theta)\coloneqq\|\hat{\mathbf F}_f\|_2^2
 =\|\mathbf P\boldsymbol\theta\|_2^2
 =\boldsymbol\theta^\top\mathbf P^\top\mathbf P\boldsymbol\theta.
\end{equation}
For the Boolean selection matrix used here, each candidate DOF is selected once, so $\mathbf P^\top\mathbf P=\mathbf I$ and $\mathcal R(\boldsymbol\theta)=\|\boldsymbol\theta\|_2^2$.
Thus $\mathcal R$ denotes the same squared load norm expressed in the new parameters. For a general mapping $\mathbf P$, it remains $\|\mathbf P\boldsymbol\theta\|_2^2$ and need not equal $\|\boldsymbol\theta\|_2^2$.

In PDE-constrained optimization, the displacement field determined by the governing equation is called the state. Here it is represented by the free-DOF displacement amplitude vector $\mathbf U\equiv\hat{\mathbf U}_f$. To distinguish this displacement from the force parameters before eliminating it as an independent unknown, define the governing residual as $\mathbf R(\mathbf U,\boldsymbol\theta)=\mathbf D_{ff}\mathbf U-\mathbf P\boldsymbol\theta$.
Using this residual and the penalty in \cref{eq:force_penalty_mapping}, the same acceleration-fitting problem can be written with an explicit equality constraint:
\begin{align}\label{eq:pdeco_general}
 \min_{\boldsymbol\theta,\mathbf U}\;\mathcal J(\boldsymbol\theta,\mathbf U)
 =\|-\omega^2\mathbf S_f\mathbf U-\hat{\mathbf a}_{\mathrm{obs}}\|_2^2
 +\alpha\mathcal R(\boldsymbol\theta)
 \quad\text{s.t.}\quad\mathbf R(\mathbf U,\boldsymbol\theta)=\mathbf0.
\end{align}
Here $\mathbf U\in\mathbb R^{N_f}$ is the free-DOF displacement amplitude, $\mathbf R$ is the vector equilibrium residual, and $\mathcal R$ is the scalar load penalty. The observations and their acceleration misfit are unchanged from \cref{eq:pdeco_objective_overview}; only the load parameterization and the explicit treatment of the displacement have changed.

For an invertible dynamic stiffness matrix $\mathbf D_{ff}$, each prescribed load $\mathbf P\boldsymbol\theta$ determines one displacement field, $\mathbf U(\boldsymbol\theta)=\mathbf D_{ff}^{-1}\mathbf P\boldsymbol\theta$. The optimization therefore searches only over the load parameters $\boldsymbol\theta$. Sparse sensors observe only part of this field: the plate example in Section~\ref{sec:7.1} has 12 measured channels and 1{,}872 free displacement DOFs. Different loads can consequently produce identical sensor readings while giving different displacements elsewhere. Restricting the candidate loaded region limits the admissible loads, and the quadratic load penalty selects a unique regularized estimate by balancing sensor misfit against load magnitude. This estimate need not equal the actual applied load. Thus a unique displacement for each given load does not imply that the measured responses uniquely determine that load.

\subsection{Reduced-space formulation}\label{sec:3.2}

The problem in \cref{eq:pdeco_general} can be approached in two ways.
The full-space (or all-at-once) approach treats $(\boldsymbol{\theta}, \mathbf{U})$ as simultaneous unknowns and enforces the PDE constraint via Lagrange multipliers \citep{biegler2003large}.
The reduced-space approach, adopted in this study, eliminates $\mathbf{U}$ by solving the PDE for each trial $\boldsymbol{\theta}$. In the reduced-space approach, the PDE constraint $\mathbf{R}(\mathbf{U}, \boldsymbol{\theta}) = \mathbf{0}$ implicitly defines $\mathbf{U}$ as a function of $\boldsymbol{\theta}$:
\begin{align}\label{eq:implicit_U}
    \mathbf{R}\bigl(\mathbf{U}(\boldsymbol{\theta}),\, \boldsymbol{\theta}\bigr) = \mathbf{0}
    \quad\Longrightarrow\quad
    \mathbf{U} = \mathbf{U}(\boldsymbol{\theta}).
\end{align}
Substituting into the objective gives the reduced objective $\widetilde{\mathcal{J}}(\boldsymbol{\theta}) \coloneqq \mathcal{J}\bigl(\boldsymbol{\theta},\, \mathbf{U}(\boldsymbol{\theta})\bigr)$, which is a function of $\boldsymbol{\theta}$ alone.
The optimization problem then reduces to the unconstrained problem $\min_{\boldsymbol{\theta}}\, \widetilde{\mathcal{J}}(\boldsymbol{\theta})$.

In this study, $\boldsymbol{\theta}$ enters only in the right-hand side $\mathbf{F}$, so the forward model $\mathbf{U}(\boldsymbol{\theta}) = \mathbf{D}_{ff}^{-1}\mathbf{F}(\boldsymbol{\theta})$ is linear in $\boldsymbol{\theta}$ because $\mathbf{D}_{ff}$ is a constant operator at the prescribed frequency and $\mathbf F(\boldsymbol\theta)=\mathbf P\boldsymbol\theta$.
The reduced objective takes the explicit form:
\begin{align}\label{eq:reduced_objective_linear}
    \widetilde{\mathcal{J}}(\boldsymbol{\theta})
    = \bigl\|-\omega^2\mathbf S_f\mathbf D_{ff}^{-1}\mathbf F(\boldsymbol\theta)-\hat{\mathbf a}_{\mathrm{obs}}\bigr\|_2^2 + \alpha\,\mathcal{R}(\boldsymbol{\theta}),
\end{align}
which is a convex quadratic in $\boldsymbol{\theta}$ for the quadratic regularizer used here. The original implementation solves it iteratively using L-BFGS with gradients computed via reverse-mode AD (see Section~\ref{sec:4.solution}).
The concrete form of \cref{eq:reduced_objective_linear} for the harmonic problem is derived in Section~\ref{sec:4.inverse} as \cref{eq:pdeco_full}.

For a boundary-restricted parameterization, the force on the candidate loaded boundary $\Gamma_N$ is parameterized via a Boolean selection matrix $\mathbf{P}$, so $\hat{\mathbf{F}}_f = \mathbf{P}\boldsymbol{\theta}$.
Each $\theta_j$ represents the nodal force at a candidate loaded DOF, and the optimization determines the force amplitudes from the sensor data.

\subsection{Gradient computation}\label{sec:3.3}

Gradient-based optimization requires the total derivative $d\widetilde{\mathcal{J}}/d\boldsymbol{\theta}$.
By the chain rule:
\begin{align}\label{eq:total_derivative}
    \frac{d\widetilde{\mathcal{J}}}{d\boldsymbol{\theta}}
    = \frac{\partial \mathcal{J}}{\partial \boldsymbol{\theta}}
    + \frac{\partial \mathcal{J}}{\partial \mathbf{U}}\,\frac{d\mathbf{U}}{d\boldsymbol{\theta}}.
\end{align}
The first term $\partial \mathcal{J}/\partial \boldsymbol{\theta}$ is the direct dependence of the objective on $\boldsymbol{\theta}$ (from the regularization term).
The second term involves the sensitivity matrix $d\mathbf{U}/d\boldsymbol{\theta} \in \mathbb{R}^{N \times N_\theta}$, which captures how the PDE solution changes with the parameter.
Computing $d\mathbf{U}/d\boldsymbol{\theta}$ explicitly (the direct method) requires solving $N_\theta$ linear systems --- one per column of the sensitivity matrix --- yielding a total cost of $N_\theta$ PDE solves.
This challenge is even greater when the force is parameterized at every free DOF, since each parameter requires a separate linear solve.

The adjoint method reduces this cost to a single additional PDE solve, regardless of $N_\theta$.
The key idea is to introduce an adjoint variable $\boldsymbol{\lambda} \in \mathbb{R}^N$ --- not a physical unknown but a computational device --- that satisfies:
\begin{align}\label{eq:adjoint_equation}
    \left(\frac{\partial \mathbf{R}}{\partial \mathbf{U}}\right)^\top \boldsymbol{\lambda} = -\left(\frac{\partial \mathcal{J}}{\partial \mathbf{U}}\right)^\top.
\end{align}
The right-hand side $\partial \mathcal{J}/\partial \mathbf{U}$ represents how the objective is sensitive to each DOF of the displacement.
Solving \cref{eq:adjoint_equation} propagates this sensitivity backward through the PDE operator, yielding $\boldsymbol{\lambda}$.
The adjoint formulation obviates the need to form $d\mathbf{U}/d\boldsymbol{\theta}$ explicitly, giving the total derivative as:
\begin{align}\label{eq:adjoint_gradient}
    \frac{d\widetilde{\mathcal{J}}}{d\boldsymbol{\theta}}
    = \frac{\partial \mathcal{J}}{\partial \boldsymbol{\theta}}
    + \boldsymbol{\lambda}^\top \frac{\partial \mathbf{R}}{\partial \boldsymbol{\theta}}.
\end{align}

Direct sensitivity calculations require one linear solve per load parameter, whereas the adjoint formulation in \cref{eq:adjoint_equation,eq:adjoint_gradient} obtains the complete gradient from one forward solve and one adjoint solve. The number of solves per objective-and-gradient evaluation is therefore independent of $N_\theta$, making the adjoint approach suitable for spatially distributed load estimation \citep{biegler2003large,oberai2003inverse,zhang2026adjoint}.

The implementation combines this adjoint formulation with automatic differentiation and GPU-accelerated linear algebra. JAX applies reverse-mode AD through the forward linear solve to evaluate the gradient, avoiding a separately coded adjoint routine. With the assembled matrices stored on a GPU, the forward solve and gradient evaluation can use device-based parallel computation; finite element assembly remains an offline CPU operation \citep{jax2018github,xue2023jaxfem}.
SciPy's host-side L-BFGS optimizer uses the resulting objective and gradient to update the load parameters without forming the full Hessian. A line search may repeat these evaluations, so two solves per evaluation does not imply two solves per iteration. Section~\ref{sec:6} describes the CPU/GPU implementation, and Section~\ref{sec:7} reports the computational costs and convergence status.

\section{A specific PDE-CO formulation for structural dynamics}\label{sec:4}

This section develops the harmonic governing equations, finite element discretization, and load and sensor models before formulating and solving the inverse problem. Figure~\ref{fig:flowchart} summarizes the computational workflow.

\subsection{Governing equations and harmonic response}\label{sec:4.governing}

Consider a structure occupying domain $\Omega \subset \mathbb{R}^3$ with boundary $\partial\Omega = \Gamma_D \cup \Gamma_N$, where $\Gamma_D$ is the clamped (fixed) boundary and $\Gamma_N$ is the boundary on which external surface traction is applied.
The displacement field $\mathbf{u}(\mathbf{x}, t)$ satisfies the linear elastodynamics initial-boundary value problem:
\begin{align}
    \rho \ddot{\mathbf{u}} - \nabla \cdot \boldsymbol{\sigma}(\mathbf{u}) &= \mathbf{b}(\mathbf{x}, t) \quad \text{in } \Omega \times (0, T], \label{eq:elastodynamics}\\ 
    \boldsymbol{\sigma}(\mathbf{u}) \cdot \mathbf{n} &= \mathbf{t}(\mathbf{x}, t) \quad \text{on } \Gamma_N, \label{eq:neumann}\\
    \mathbf{u} &= \mathbf{0} \quad \text{on } \Gamma_D, \label{eq:dirichlet}
\end{align}
where $\rho$ is the mass density, $\boldsymbol{\sigma}$ is the Cauchy stress tensor, $\mathbf{b}$ is the body force per unit volume, $\mathbf{n}$ is the outward unit normal on $\Gamma_N$, and $\mathbf{t}$ is the surface traction.
The constitutive relation for linear isotropic elasticity is:
\begin{align}\label{eq:constitutive}
    \boldsymbol{\sigma} = \lambda\,\mathrm{tr}(\boldsymbol{\varepsilon})\,\mathbf{I} + 2\mu\,\boldsymbol{\varepsilon},
\end{align}
where $\lambda = E\nu/\bigl((1+\nu)(1-2\nu)\bigr)$ and $\mu = E/\bigl(2(1+\nu)\bigr)$ are the Lam\'{e} parameters, $E$ is Young's modulus, $\nu$ is Poisson's ratio, and the small strain tensor is $\boldsymbol{\varepsilon} = \frac{1}{2}(\nabla\mathbf{u} + \nabla\mathbf{u}^\top)$.

The present study focuses on the case of harmonic (single-frequency) excitation, which enables an algebraic reformulation that avoids time integration entirely.
Under harmonic excitation at circular frequency $\omega$, all time-dependent quantities are expressed as:
\begin{align}\label{eq:harmonic_form}
    \mathbf{u}(\mathbf{x}, t) = \hat{\mathbf{u}}(\mathbf{x})\,e^{i\omega t}, \quad
    \mathbf{t}(\mathbf{x}, t) = \hat{\mathbf{t}}(\mathbf{x})\,e^{i\omega t}, \quad
    \mathbf{b}(\mathbf{x}, t) = \hat{\mathbf{b}}(\mathbf{x})\,e^{i\omega t},
\end{align}
where the hatted quantities are the frequency-domain complex amplitudes, following the convention introduced in Section~\ref{sec:1}.
The frequency $\omega$ is assumed known --- it can be identified from the frequency content of sensor time histories (e.g., via FFT) or is prescribed in controlled laboratory tests.

Substituting \cref{eq:harmonic_form} into \cref{eq:elastodynamics} and canceling the common factor $e^{i\omega t}$, the time derivative becomes a multiplication by $-\omega^2$:
\begin{align}\label{eq:harmonic_substitution}
    \rho\,\ddot{\mathbf{u}} = \rho\,(i\omega)^2\,\hat{\mathbf{u}}\,e^{i\omega t} = -\omega^2\rho\,\hat{\mathbf{u}}\,e^{i\omega t}.
\end{align}
The governing PDE in \cref{eq:elastodynamics} thus reduces to the frequency-domain form:
\begin{align}\label{eq:harmonic_pde}
    -\omega^2 \rho\, \hat{\mathbf{u}} - \nabla \cdot \boldsymbol{\sigma}(\hat{\mathbf{u}}) = \hat{\mathbf{b}} \quad \text{in } \Omega.
\end{align}
Under linear steady-state assumptions, a frequency-by-frequency extension can decompose sensor time histories via FFT, solve at each frequency $\omega_k$ with $\mathbf{D}_{ff}(\omega_k)$, and superpose the results via the inverse fast Fourier transform (IFFT). Finite-record transients, frequency-dependent conditioning and noise regularization require additional treatment; a broadband reconstruction is not validated here. The single-frequency formulation supplies the building block for such an extension.

\subsection{Finite element discretization and boundary conditions}\label{sec:4.discretization}

\cref{eq:harmonic_pde} is discretized using the standard Galerkin finite element method.
The displacement field is approximated as $\mathbf{u}^h(\mathbf{x}) = \sum_{I=1}^{N_{\mathrm{node}}} N_I(\mathbf{x})\,\mathbf{d}_I$, where $N_I$ is the shape function associated with node $I$ and $\mathbf{d}_I \in \mathbb{R}^3$ is the nodal displacement vector.

Multiplying \cref{eq:elastodynamics} by a test function $\mathbf{v}^h$, integrating over $\Omega$, applying integration by parts, and substituting the FE approximation yields the semi-discrete equation of motion:
\begin{align}\label{eq:semidiscrete}
    \mathbf{M}\ddot{\mathbf{U}}(t) + \mathbf{K}\mathbf{U}(t) = \mathbf{R}(t),
\end{align}
where $\mathbf{U}(t) \in \mathbb{R}^N$ is the global nodal displacement vector, $\mathbf{M} \in \mathbb{R}^{N \times N}$ is the consistent mass matrix, $\mathbf{K} \in \mathbb{R}^{N \times N}$ is the stiffness matrix, and $\mathbf{R}(t) \in \mathbb{R}^N$ is the consistent nodal force vector assembled from body forces and surface tractions.
Damping is omitted in the original three-case benchmark. The additional noise-sensitivity tests include mass-proportional damping, with damping matrix $\mathbf{C}_d$, as specified below.

For harmonic excitation, substituting $\mathbf{U}(t) = \hat{\mathbf{U}}\,e^{i\omega t}$ and $\mathbf{R}(t) = \hat{\mathbf{R}}\,e^{i\omega t}$ into \cref{eq:semidiscrete} and defining the dynamic stiffness matrix $\mathbf{D}(\omega) \coloneqq \mathbf{K} - \omega^2 \mathbf{M}$:
\begin{align}\label{eq:dyn_system}
    \mathbf{D}(\omega)\, \hat{\mathbf{U}} = \hat{\mathbf{R}}.
\end{align}
The dynamic stiffness $\mathbf{D}(\omega)$ combines the elastic restoring force ($\mathbf{K}$) and the inertial reaction ($-\omega^2\mathbf{M}$) into a single operator.
At low frequencies ($\omega \to 0$), $\mathbf{D} \approx \mathbf{K}$ and the problem reduces to the quasi-static regime.
At a natural frequency $\omega = \omega_n$, $\mathbf{D}$ becomes singular (resonance).
The excitation frequency $\omega$ must therefore be chosen away from any natural frequency.

The global DOFs are partitioned into two disjoint sets based on their boundary conditions:
\begin{align}\label{eq:dof_partition}
    \hat{\mathbf{U}} = \begin{bmatrix} \hat{\mathbf{U}}_f \\ \hat{\mathbf{U}}_c \end{bmatrix}, \quad
    \hat{\mathbf{R}} = \begin{bmatrix} \hat{\mathbf{R}}_f \\ \hat{\mathbf{R}}_c \end{bmatrix},
\end{align}
where subscript $f$ denotes free DOFs (unknown displacements) and subscript $c$ denotes clamped DOFs (prescribed as $\hat{\mathbf{U}}_c = \mathbf{0}$ on $\Gamma_D$).

The dynamic stiffness system in \cref{eq:dyn_system} partitions as:
\begin{align}\label{eq:block_system}
    \begin{bmatrix}
        \mathbf{D}_{ff} & \mathbf{D}_{fc} \\
        \mathbf{D}_{cf} & \mathbf{D}_{cc}
    \end{bmatrix}
    \begin{bmatrix}
        \hat{\mathbf{U}}_f \\ \mathbf{0}
    \end{bmatrix}
    =
    \begin{bmatrix}
        \hat{\mathbf{F}}_f \\ \hat{\mathbf{R}}_c
    \end{bmatrix},
\end{align}
where $\hat{\mathbf{F}}_f$ is the external force on the free DOFs (including the unknown traction on $\Gamma_N$), and $\hat{\mathbf{R}}_c$ is the reaction force at the clamped boundary (not needed for the present problem).
Since $\hat{\mathbf{U}}_c = \mathbf{0}$, the first block row gives:
\begin{align}\label{eq:free_system}
    \mathbf{D}_{ff}\, \hat{\mathbf{U}}_f = \hat{\mathbf{F}}_f,
\end{align}
where $\mathbf{D}_{ff} = \mathbf{K}_{ff} - \omega^2\mathbf{M}_{ff} \in \mathbb{R}^{N_f \times N_f}$ is the free-DOF dynamic stiffness matrix.
This is the governing equation of the present inverse problem: given observations related to $\hat{\mathbf{U}}_f$, the objective is to recover the right-hand side $\hat{\mathbf{F}}_f$.

For the additional damped studies, the same free-DOF equation is extended to
\begin{equation}\label{eq:forward}
 \mathbf D_{ff}(\omega)\hat{\mathbf U}_f=\mathbf P\boldsymbol\theta,
 \qquad \mathbf D_{ff}(\omega)=\mathbf K_{ff}+i\omega\mathbf C_{d,ff}-\omega^2\mathbf M_{ff}.
\end{equation}
Here $\mathbf C_{d,ff}=c_M\mathbf M_{ff}$ is the mass-proportional damping matrix, and the amplitudes can be complex. Setting $c_M=0$ recovers the undamped formulation above. The original real-valued derivations below apply directly to the undamped benchmark; their complex least-squares counterparts additionally require complex conjugation, denoted by an overline.

\subsection{Force parameterization and sensor observations}\label{sec:4.measurements}

The force vector $\hat{\mathbf{F}}_f \in \mathbb{R}^{N_f}$ has $N_f$ components --- one for each free DOF.
Although the loaded boundary $\Gamma_N$ may be known from engineering context (e.g., an inner pressure surface or a contact face), the distribution of force across $\Gamma_N$ is unknown and must be recovered from sensor data.

To this end, a force parameterization is introduced that maps a reduced set of unknowns on $\Gamma_N$ to the full force vector.
Let $\boldsymbol{\theta} \in \mathbb{R}^{N_\theta}$ be the vector of unknown force amplitudes at the $N_\theta$ candidate DOFs.
The relationship between $\boldsymbol{\theta}$ and the full force vector is:
\begin{align}\label{eq:force_param}
    \hat{\mathbf{F}}_f = \mathbf{P}\boldsymbol{\theta},
\end{align}
where $\mathbf{P} \in \mathbb{R}^{N_f \times N_\theta}$ is the force mapping matrix (also called the load distribution matrix).
Together with the sensor extraction matrix $\mathbf{S}_f$, $\mathbf{P}$ defines the input--output pipeline of PDE-CO: $\mathbf{P}$ maps $\boldsymbol{\theta}$ into the full DOF space (input selection), while $\mathbf{S}_f$ extracts sensor DOFs from the solution (output selection).
When the loaded boundary $\Gamma_N$ is known a priori, $\mathbf{P}$ is constructed as a Boolean selection matrix that picks out the DOFs on $\Gamma_N$:
\begin{align}\label{eq:P_boolean}
    P_{ij} = \begin{cases} 1 & \text{if free DOF $i$ corresponds to loaded DOF $j$}, \\ 0 & \text{otherwise}. \end{cases}
\end{align}
Each $\theta_j$ then represents a nodal force value (in N) at a known loaded DOF, and $N_\theta$ equals the number of loaded DOFs on $\Gamma_N$.
When $\Gamma_N$ is a localized region, $N_\theta$ can be significantly smaller than $N_f$; in general, $N_\theta$ typically remains larger than $N_s$, so the inverse problem is underdetermined from a purely data-driven perspective.
For each trial $\boldsymbol{\theta}$, the governing equation determines the full displacement field $\mathbf{U}(\boldsymbol{\theta})$, enforcing consistency with the FE model across the domain. The data-to-force map can still be rank deficient. Positive Tikhonov regularization selects a unique regularized estimate; the minimum-norm least-squares solution is obtained in the limit $\alpha\to0^+$. Neither statement guarantees that the estimate equals the physical force.

Acceleration measurements are widely used for response and force estimation in structural dynamics \citep{lourens2012augmented,kullaa2016virtual,iliopoulos2016modal,maes2016dynamic,tarpo2020expansion}.
Under harmonic excitation, the acceleration at a point is related to the displacement by:
\begin{align}\label{eq:accel_relation}
    \mathbf{a}(\mathbf{x}, t) = \ddot{\mathbf{u}}(\mathbf{x}, t) = -\omega^2\,\hat{\mathbf{u}}(\mathbf{x})\,e^{i\omega t}.
\end{align}
Thus the acceleration amplitude is simply $\hat{\mathbf{a}} = -\omega^2\,\hat{\mathbf{u}}$: a scalar multiple of the displacement amplitude.
As noted in Section~\ref{sec:4.governing}, the single-frequency formulation extends naturally to broadband excitation via FFT-based frequency decomposition and superposition.

Suppose $N_{\mathrm{sensor}}$ accelerometer nodes are placed at free locations, including observable exterior surfaces, providing $N_s = 3 \times N_{\mathrm{sensor}}$ scalar observations (three components per node).
The observed acceleration amplitudes are collected in a vector $\hat{\mathbf{a}}_{\mathrm{obs}} \in \mathbb{R}^{N_s}$. To relate $\hat{\mathbf{a}}_{\mathrm{obs}}$ to the full free-DOF displacement vector $\hat{\mathbf{U}}_f$, a sensor selection matrix $\mathbf{S}_f \in \mathbb{R}^{N_s \times N_f}$ is introduced.
This is a Boolean matrix that extracts sensor DOFs from the free DOF vector:
\begin{align}\label{eq:selection_matrix}
    (S_f)_{ij} = \begin{cases} 1 & \text{if sensor DOF $i$ corresponds to free DOF $j$}, \\ 0 & \text{otherwise}. \end{cases}
\end{align}
Each row of $\mathbf{S}_f$ contains exactly one nonzero entry.
The observation equation then reads:
\begin{align}\label{eq:accel_observation}
    \hat{\mathbf{a}}_{\mathrm{obs}} = -\omega^2\, \mathbf{S}_f\, \hat{\mathbf{U}}_f.
\end{align}

\subsection{Reduced harmonic inverse problem}\label{sec:4.inverse}
Combining the force mapping, free-DOF equilibrium and acceleration observation gives
\begin{equation}\label{eq:forward_chain}
 \hat{\mathbf U}_f(\boldsymbol\theta)=\mathbf D_{ff}^{-1}\mathbf P\boldsymbol\theta,\quad \hat{\mathbf a}_{\mathrm{pred}}=\mathbf A\boldsymbol\theta,\quad \mathbf A=-\omega^2\mathbf S_f\mathbf D_{ff}^{-1}\mathbf P.
\end{equation}
The assembled matrices, force mapping and sensor layout are prescribed; the excitation frequency is known and the observed acceleration is the input to reconstruction. The force parameters are the optimization variables, and the governing equation determines the displacement for each candidate load. The reduced problem is
\begin{equation}\label{eq:pdeco_full}
 \min_{\boldsymbol\theta}\;\mathcal J(\boldsymbol\theta)
 =\|\mathbf A\boldsymbol\theta-\hat{\mathbf a}_{\mathrm{obs}}\|_2^2
 +\alpha\|\boldsymbol\theta\|_2^2,\qquad\alpha>0.
\end{equation}
For $\alpha>0$, the objective is strictly convex and has a unique minimizer; this uniqueness does not establish physical load identifiability.

\subsection{AD-based solution and full-field reconstruction}\label{sec:4.solution}
The original iterative implementation evaluates \cref{eq:pdeco_full} without explicitly forming $\mathbf A$. For the real undamped benchmark, the forward solve gives $\hat{\mathbf U}_f$ and the sensor residual $\mathbf r=-\omega^2\mathbf S_f\hat{\mathbf U}_f-\hat{\mathbf a}_{\mathrm{obs}}$. The adjoint variable here is minus one half of the general Lagrange multiplier in \cref{eq:adjoint_equation}, so the factor of two is retained in the gradient. The adjoint and gradient are
\begin{align}
 \mathbf D_{ff}^{\top}\boldsymbol\lambda&=-\omega^2\mathbf S_f^{\top}\mathbf r,\label{eq:harmonic_adjoint}\\
 \nabla_{\boldsymbol\theta}\mathcal J&=2\mathbf P^{\top}\boldsymbol\lambda+2\alpha\boldsymbol\theta.\label{eq:gradient_adjoint}
\end{align}
For complex amplitudes, the adjoint solve uses $\overline{\mathbf D_{ff}}^{\top}$ in place of $\mathbf D_{ff}^{\top}$. Reverse-mode AD through the JAX linear solve evaluates this adjoint gradient automatically \citep{xue2023jaxfem,jax2018github}. SciPy's L-BFGS-B optimizer uses the gradient to update the force parameters; no bounds are imposed in the reported runs \citep{virtanen2020scipy}.

One objective-and-gradient evaluation requires two linear solves, independent of the number of load parameters. A cached factorization can serve repeated evaluations, but this reuse must be implemented explicitly. Section~\ref{sec:6.2} distinguishes the original dense-solve code from the factorization-reuse algorithm. Conditioning and line searches affect the iteration cost; reaching the iteration budget is distinguished from convergence.

Once the optimal load $\boldsymbol{\theta}^*$ is obtained, the complete structural response is recovered by a single forward solve:
\begin{equation}\label{eq:reconstruct}
    \hat{\mathbf{U}}_f^* = \mathbf{D}_{ff}^{-1}\mathbf{P}\boldsymbol{\theta}^*, \qquad \hat{\mathbf{a}}_{\mathrm{full}}^* = -\omega^2\,\hat{\mathbf{U}}_f^*.
\end{equation}

The output of PDE-CO is the full displacement and acceleration fields over the entire domain $\Omega$, not just at sensor locations.
The recovered load $\mathbf P\boldsymbol\theta^*$ and its associated displacement are assessed separately, since their reconstruction errors need not be comparable.

\section{Computational implementation}\label{sec:6}

\begin{figure}[t]
\centering
\includegraphics[width=0.85\linewidth]{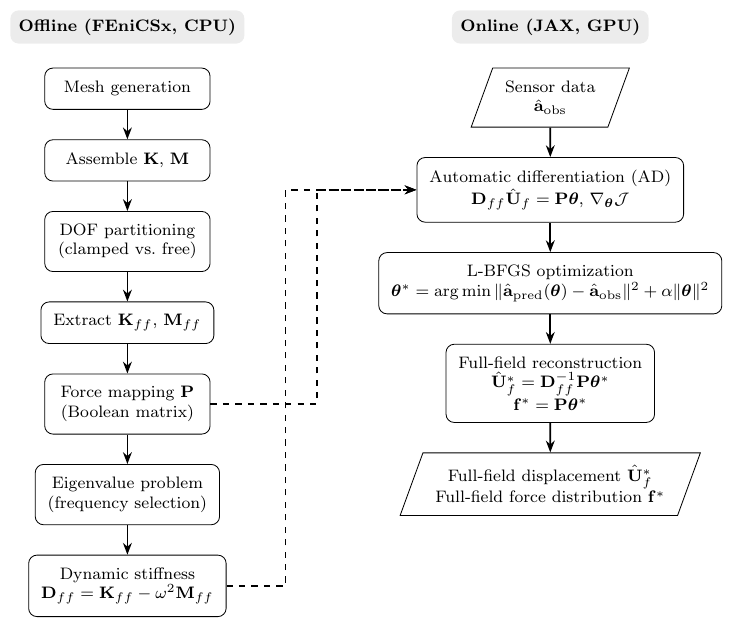}
\caption{Original iterative PDE-CO workflow for the undamped harmonic benchmark. FEniCSx assembles the free-DOF matrices and load map; the JAX branch evaluates the objective and its gradient and updates the force using L-BFGS. The separate sparse sensor-space reference solution and the damped uncertainty studies are described in the text. The CPU/GPU labels identify the implementation stages and are not an end-to-end performance comparison.}\label{fig:flowchart}
\end{figure}

Figure~\ref{fig:flowchart} illustrates the overall pipeline, which consists of two stages.
The offline stage is executed on a CPU using FEniCSx and proceeds as follows:
(i)~a finite element mesh is generated for the target structure;
(ii)~the global stiffness and mass matrices $\mathbf{K}$ and $\mathbf{M}$ are assembled;
(iii)~the degrees of freedom are partitioned into clamped (Dirichlet) and free sets;
(iv)~the free-DOF submatrices $\mathbf{K}_{ff}$ and $\mathbf{M}_{ff}$ are extracted;
(v)~the Boolean force mapping matrix $\mathbf{P}$ is constructed to map the reduced force parameter $\boldsymbol{\theta}$ to the full free-DOF force vector;
(vi)~an eigenvalue problem is solved to select the excitation frequency $\omega$; and
(vii)~the frequency-domain dynamic stiffness matrix $\mathbf{D}_{ff} = \mathbf{K}_{ff} - \omega^2\mathbf{M}_{ff}$ is formed.
The JAX implementation can execute forward and gradient evaluations on a GPU, while SciPy controls the optimizer on the host. The saved original summaries do not record the device used for each run.
For the GPU configuration illustrated in Figure~\ref{fig:flowchart}, the offline-computed matrices $\mathbf{D}_{ff}$ and $\mathbf{P}$ are transferred to the device. Given sensor observations $\hat{\mathbf{a}}_{\mathrm{obs}}$, the following operations are performed:
(i)~the forward problem $\mathbf{D}_{ff}\hat{\mathbf{U}}_f = \mathbf{P}\boldsymbol{\theta}$ and the gradient $\nabla_{\boldsymbol{\theta}}\mathcal{J}$ are evaluated via automatic differentiation;
(ii)~the L-BFGS optimizer iteratively updates $\boldsymbol{\theta}$ to minimize $\|\hat{\mathbf{a}}_{\mathrm{pred}}(\boldsymbol{\theta}) - \hat{\mathbf{a}}_{\mathrm{obs}}\|^2 + \alpha\|\boldsymbol{\theta}\|^2$; and
(iii)~at termination, the full-field displacement $\hat{\mathbf{U}}_f^* = \mathbf{D}_{ff}^{-1}\mathbf{P}\boldsymbol{\theta}^*$ and the full-field force distribution $\mathbf{f}^* = \mathbf{P}\boldsymbol{\theta}^*$ are reconstructed.

\subsection{Finite element implementation (offline stage)}\label{sec:6.1}

The spatial discretization follows the standard Galerkin finite element method using FEniCSx \citep{baratta2023dolfinx}.
The displacement field is interpolated using continuous Lagrange shape functions as described in Section~\ref{sec:4.discretization}.
The trial and test function spaces are defined as
\begin{align}
    \mathcal{S}^h &= \bigl\{\mathbf{u}^h \in [H^1(\Omega)]^{n_{\mathrm{dim}}} \;\big|\; \mathbf{u}^h = \bar{\mathbf{u}} \text{ on } \Gamma_D \bigr\}, \\
    \mathcal{V}^h &= \bigl\{\mathbf{v}^h \in [H^1(\Omega)]^{n_{\mathrm{dim}}} \;\big|\; \mathbf{v}^h = \mathbf{0} \text{ on } \Gamma_D \bigr\},
\end{align}
where $n_{\mathrm{dim}}$ denotes the spatial dimension.
The body-force density is set to zero in the present benchmark. The corresponding weak form of the governing equation in \cref{eq:elastodynamics} is stated as follows: find $\mathbf{u}^h \in \mathcal{S}^h$ such that
\begin{align}\label{eq:weak_form}
    \int_\Omega \boldsymbol{\sigma}(\mathbf{u}^h) : \boldsymbol{\varepsilon}(\mathbf{v}^h)\,\mathrm{d}V
    + \int_\Omega \rho\,\ddot{\mathbf{u}}^h \cdot \mathbf{v}^h\,\mathrm{d}V
    = \int_{\Gamma_N} \mathbf{t} \cdot \mathbf{v}^h\,\mathrm{d}S
    \qquad \forall\,\mathbf{v}^h \in \mathcal{V}^h.
\end{align}
The small strain tensor is discretized as
\begin{align}\label{eq:strain_interp}
    \boldsymbol{\varepsilon}^h = \sum_{a=1}^{n_{\mathrm{node}}} \mathbf{B}_a\,\mathbf{u}_a,
\end{align}
where $\mathbf{B}_a = \nabla^s N_a$ is the symmetric gradient of the shape function associated with node $a$.
Substituting the FE interpolation into \cref{eq:weak_form}, the global stiffness matrix $\mathbf{K}$, consistent mass matrix $\mathbf{M}$, and consistent nodal force vector $\mathbf{R}$ are assembled from element-level contributions:
\begin{align}
    \mathbf K_{ab} &= \int_\Omega \mathbf{B}_a^\top\,\mathbb{C}\,\mathbf{B}_b\,\mathrm{d}V, \label{eq:K_assembly}\\
    \mathbf M_{ab} &= \int_\Omega \rho\,N_a\,N_b\,\mathbf{I}\,\mathrm{d}V, \label{eq:M_assembly}\\
    \mathbf R_a &= \int_{\Gamma_N} N_a\,\mathbf{t}\,\mathrm{d}S, \label{eq:R_assembly}
\end{align}
where $\mathbb{C}$ is the fourth-order elasticity tensor corresponding to the linear isotropic constitutive relation in \cref{eq:constitutive}, and $\rho$ is the mass density.
The surface traction $\mathbf{t}$ motivates the inverse load model, while the original computations parameterize nodal force amplitudes as $\hat{\mathbf{F}}_f = \mathbf{P}\boldsymbol{\theta}$ via \cref{eq:P_boolean}, or use the unrestricted identity map. A Boolean nodal-force selector is not itself a continuous traction basis or the integration in \cref{eq:R_assembly}, and a Euclidean penalty on nodal forces is discretisation dependent rather than a mesh-invariant traction prior.

Applying the DOF partition~(\cref{eq:dof_partition}, Section~\ref{sec:4.discretization}) to the global stiffness and mass matrices yields the block structure:
\begin{align}\label{eq:block_KM}
    \mathbf{K} = \begin{bmatrix} \mathbf{K}_{ff} & \mathbf{K}_{fc} \\ \mathbf{K}_{cf} & \mathbf{K}_{cc} \end{bmatrix}, \qquad
    \mathbf{M} = \begin{bmatrix} \mathbf{M}_{ff} & \mathbf{M}_{fc} \\ \mathbf{M}_{cf} & \mathbf{M}_{cc} \end{bmatrix}.
\end{align}
Since the homogeneous Dirichlet condition $\hat{\mathbf{U}}_c = \mathbf{0}$ eliminates the coupling terms $\mathbf{K}_{fc}\hat{\mathbf{U}}_c$ and $\mathbf{M}_{fc}\hat{\mathbf{U}}_c$ from the governing equation in \cref{eq:free_system}, only the free-DOF sub-matrices $\mathbf{K}_{ff}$, $\mathbf{M}_{ff} \in \mathbb{R}^{N_f \times N_f}$ are required.

For the mode-based approach (Section~\ref{sec:2}), which will be compared with the proposed method in the numerical examples (Section~\ref{sec:7}), the generalized eigenvalue problem
\begin{align}\label{eq:eigenvalue}
    \mathbf{K}_{ff}\boldsymbol{\phi}_n = \omega_n^2\,\mathbf{M}_{ff}\boldsymbol{\phi}_n
\end{align}
is solved to obtain the natural frequencies $\omega_n$ and mode shapes $\boldsymbol{\phi}_n$.
The dynamic stiffness matrix is then formed at the excitation frequency $\omega$:
\begin{align}\label{eq:Dff_assembly}
    \mathbf{D}_{ff} = \mathbf{K}_{ff} - \omega^2\mathbf{M}_{ff}.
\end{align}

\subsection{Automatic differentiation using JAX (online stage)}\label{sec:6.2}

In the online stage, JAX differentiates through the forward linear solve, and SciPy's L-BFGS-B optimizer uses the resulting gradient to update the load parameters in \cref{eq:pdeco_full}. Forward and gradient evaluations can run on a GPU, while the optimizer runs on the CPU \citep{jax2018github,virtanen2020scipy}.

Algorithm~\ref{alg:online} summarizes the procedure with explicit factorization reuse. The original implementation uses dense \texttt{jnp.linalg.solve} calls without established reuse across iterations; the sparse direct reference reuses its factors. At termination, \cref{eq:reconstruct} gives the full-field response.

\begin{algorithm}[t]
\caption{PDE-CO for full-field inference (harmonic, linear, AD-based; factorization-reuse form)}\label{alg:online}
\begin{algorithmic}[1]
\Require $\mathbf{D}_{ff}$, $\mathbf{P}$, $\mathbf{S}_f$; observed $\hat{\mathbf{a}}_{\mathrm{obs}}$; frequency $\omega$; regularization $\alpha$; max iterations $k$
\Ensure Load estimate $\boldsymbol{\theta}^*$; recovered field $\hat{\mathbf{U}}_f^*$; termination status
\State Factorize $\mathbf{D}_{ff} = \mathbf{L}\mathbf{U}$ \Comment{One-time LU factorization}
\State $\boldsymbol{\theta} \gets \mathbf{0}$ \Comment{Initial guess}
\For{$i = 1, \ldots, k$}
    \State $\hat{\mathbf{U}}_f \gets \mathbf{D}_{ff}^{-1}\mathbf{P}\boldsymbol{\theta}$ \Comment{Forward solve (back-substitution)}
    \State $\hat{\mathbf{a}}_{\mathrm{pred}} \gets -\omega^2\,\mathbf{S}_f\hat{\mathbf{U}}_f$; \quad $\mathbf{r} \gets \hat{\mathbf{a}}_{\mathrm{pred}} - \hat{\mathbf{a}}_{\mathrm{obs}}$
    \State $\boldsymbol{\lambda} \gets \mathbf{D}_{ff}^{-1}(-\omega^2\,\mathbf{S}_f^\top\mathbf{r})$ \Comment{Adjoint solve (back-substitution)}
    \State $\mathbf{g} \gets 2\mathbf{P}^\top\boldsymbol{\lambda} + 2\alpha\boldsymbol{\theta}$ \Comment{Gradient}
    \State $\boldsymbol{\theta} \gets \mathrm{L\text{-}BFGS\text{-}update}(\boldsymbol{\theta}, \mathbf{g})$ \Comment{Line search may repeat evaluations}
    \If{the optimizer stopping criterion is satisfied}
        \State \textbf{break}
    \EndIf
\EndFor
\State $\boldsymbol{\theta}^* \gets \boldsymbol{\theta}$
\State $\hat{\mathbf{U}}_f^* \gets \mathbf{D}_{ff}^{-1}\mathbf{P}\boldsymbol{\theta}^*$ \Comment{Full-field reconstruction}
\end{algorithmic}
\end{algorithm}

\section{Numerical examples}\label{sec:7}

Three benchmark problems of progressively increasing complexity are presented to evaluate the proposed framework using synthetic measurements:
(1)~a cantilever rectangular plate subjected to a Gaussian transverse nodal load on the free-end face;
(2)~a $90^\circ$ elbow pipe subjected to uniform nodal loading on the free-end cross-section;
and (3)~a reactor pressure vessel (RPV) subjected to a localized Gaussian harmonic nodal load on the inner surface.
The cases progress in both geometric complexity and load configuration, allowing the displacement and force reconstructions to be assessed under different spatial loading conditions.
The original three examples assume a linear isotropic elastic material (structural steel), undamped harmonic steady-state excitation at $\omega = \omega_1/2$ (sub-resonant regime), and benchmark PDE-CO against the mode-based approach (Section~\ref{sec:2}).
For the mode-based approach, the retained number of modes, $N_d$, is 11 for the plate, 23 for the pipe and 50 for the RPV.
Reconstruction accuracy is measured by the relative displacement and force errors:
\begin{equation}
 \epsilon_u=\frac{\|\hat{\mathbf U}_f^{\mathrm{rec}}-\hat{\mathbf U}_f^{\mathrm{true}}\|_2}{\|\hat{\mathbf U}_f^{\mathrm{true}}\|_2},\qquad
 \epsilon_f=\frac{\|\hat{\mathbf F}_f^{\mathrm{rec}}-\hat{\mathbf F}_f^{\mathrm{true}}\|_2}{\|\hat{\mathbf F}_f^{\mathrm{true}}\|_2}.
\end{equation}
Additional tests assess measurement-noise sensitivity and output ambiguity.

\begin{table}[htbp]
\centering
\caption{Summary of numerical results for the three examples.
$N_f$: free DOFs; $N_\theta$: force parameter dimension; $N_{\rm sensor}$: number of sensors;
$\epsilon_u$: relative displacement error; Iters: L-BFGS iterations;
$t_{\rm opt}$: optimization wall time recorded for the original JAX runs, excluding JIT compilation. Pipe and RPV reached their iteration limits without reported convergence. These historical timings are not an end-to-end CPU/GPU comparison.}\label{tab:results_summary}
\footnotesize
\begin{tabular}{lrrrrrrr}
\toprule
 & \multirow{2}{*}{$N_f$} & \multirow{2}{*}{$N_\theta$} & \multirow{2}{*}{$N_{\rm sensor}$} & \multicolumn{2}{c}{$\epsilon_u$ [\%]} & \multirow{2}{*}{Iters} & \multirow{2}{*}{$t_{\rm opt}$ [s]} \\
\cmidrule(lr){5-6}
 &  &  &  & PDE-CO & \shortstack{Mode-based\\approach} &  &  \\
\midrule
Plate & 1{,}872 & 1{,}872 & 4 & 0.45 & 7.69 & 11 & 0.53 \\
Pipe  & 11{,}160 & 11{,}160 & 8 & 0.28 & 1.11 & 200 & 615 \\
RPV   & 7{,}206 & 2{,}850 & 36 & 0.885 & 1.86 & 2{,}000 & 15{,}294 \\
\bottomrule
\end{tabular}
\end{table}

\subsection{Rectangular plate}\label{sec:7.1}

\begin{figure}[htbp]
\centering
\includegraphics[width=0.95\linewidth]{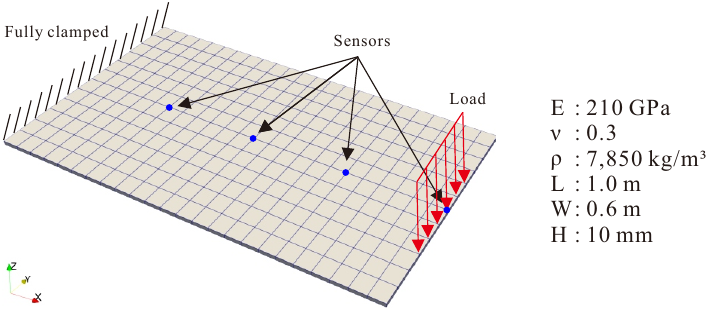}
\caption{Cantilever plate geometry, material properties, clamped face and four acceleration sensors. The red arrows indicate the loaded tip face and transverse direction schematically. The numerical reference is a Gaussian $z$-directed nodal force across that face; the arrow distribution does not specify a uniform load or its harmonic phase.}\label{fig:plate_overview}
\end{figure}

The first example is a cantilever rectangular plate
($L_x \times L_y \times L_z = 1.0 \times 0.6 \times 0.01$~m,
$E = 210$~GPa, $\nu = 0.3$, $\rho = 7{,}850$~kg/m\textsuperscript{3}).
The plate is fully clamped at $x = 0$ (zero displacement on all faces at $x = 0$) and free on the remaining five faces.
A Gaussian $z$-direction nodal force is applied on the free-end face ($x = L_x$), represented schematically by red arrows in Figure~\ref{fig:plate_overview}. The saved reference amplitudes are positive in the computational $z$ direction; the arrows indicate the loaded region and transverse direction, without fixing the harmonic phase.
The force is parameterized by $\boldsymbol{\theta} \in \mathbb{R}^{1{,}872}$ covering all free DOFs ($N_\theta=1{,}872$, $\mathbf{P}=\mathbf{I}$) in the original result. A separate surface-selective parameterization with 78 tip-face components is assessed below; it is not the load space used in the original field plots.
Four acceleration sensors (12 DOFs) are placed on the top surface.
The FE mesh consists of 288 trilinear hexahedral elements (650 nodes, $N_f = 1{,}872$ free DOFs).
The excitation frequency is set to $\omega = 74.3$~rad/s (${\approx}11.8$~Hz, i.e., $\omega_1/2$), and the Tikhonov regularization parameter is $\alpha = 10^{-6}$.

\begin{figure}[htbp]
\centering
\includegraphics[width=0.9\linewidth]{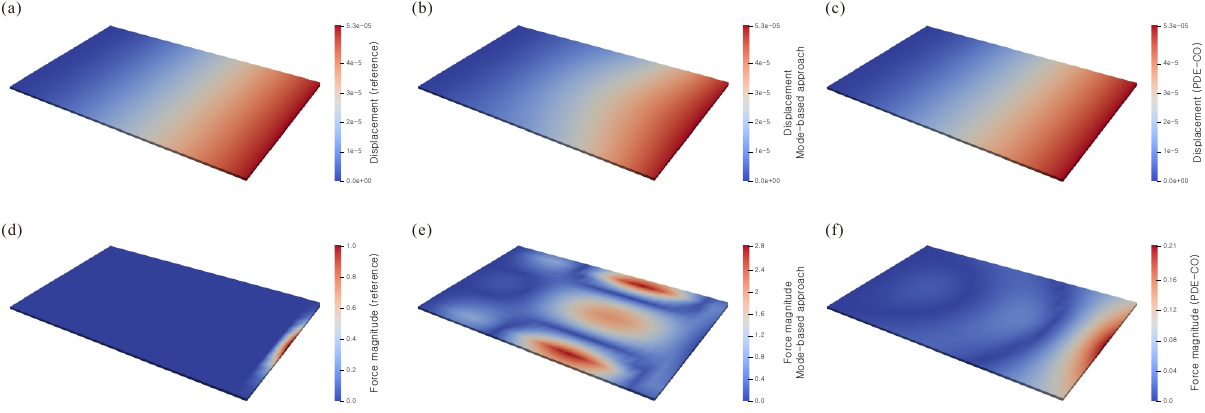}
\caption{Cantilever plate: original iterative reconstruction benchmark. Upper row: displacement magnitude (m) for (a) reference, (b) the mode-based approach and (c) PDE-CO. Lower row: corresponding nodal-force magnitudes (N). The displayed PDE-CO run estimates forces at all 1{,}872 free DOFs. Force panels use their displayed individual scales; visual localisation does not imply accurate force amplitudes. The full nodal-force error of this PDE-CO result is 83.50\%.}\label{fig:plate_convergence}
\end{figure}

Figure~\ref{fig:plate_convergence} presents the full-field comparison of the recovered displacement and force distributions (six panels).
All field quantities are visualized on the undeformed mesh geometry to enable direct visual comparison between methods.
The upper row shows the displacement magnitude field --- (a)~reference, (b)~the mode-based approach, (c)~PDE-CO.
Both methods reproduce the gross bending pattern (large displacement at the free end, zero at the clamp), but PDE-CO matches the reference field more closely.
The lower row shows the recovered force magnitude distribution over the plate surface --- (d)~reference, (e)~the mode-based approach, (f)~PDE-CO.
The reference force is sharply concentrated on the free-end face.
The force estimate from the mode-based approach (e) contains multiple hotspots scattered across the interior of the plate, away from the true loading region.
The PDE-CO force estimate (f) shows larger magnitudes near the free-end region, but its full nodal-force error is $\epsilon_f=83.50\%$; the displayed localization does not establish accurate force recovery.

\begin{figure}[htbp]
\centering
\includegraphics[width=0.9\linewidth]{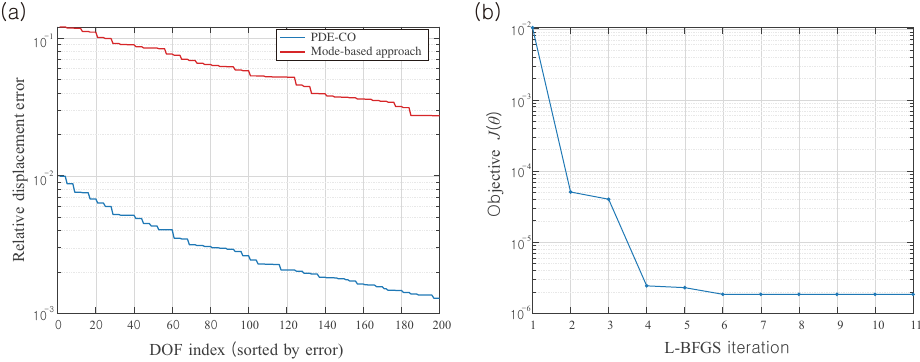}
\caption{Cantilever plate: original iterative benchmark diagnostics. (a) Largest 200 component errors $|\hat U_{f,i}^{\rm rec}-\hat U_{f,i}^{\rm true}|/\max_k|\hat U_{f,k}^{\rm true}|$, ranked independently for each method; the abscissa is error rank rather than a shared physical DOF. The full-vector displacement errors are 0.45\% for PDE-CO and 7.69\% for the mode-based approach. (b) Recorded objective history for the 11-iteration run, which met its optimiser termination criterion.}\label{fig:plate_disp_error}
\end{figure}

Figure~\ref{fig:plate_disp_error} presents (a)~relative component displacement errors ranked independently by magnitude (top 200) and (b)~the L-BFGS objective history.
The L-BFGS optimizer met its termination criterion in 11~iterations ($\mathcal{J} \approx 10^{-6}$).
As reported in Table~\ref{tab:results_summary} (row 1), PDE-CO achieves $\epsilon_u = 0.45\%$ while the mode-based approach with 4 sensors yields $\epsilon_u = 7.69\%$.
While the mode-based approach ($7.69\%$) provides a reasonable displacement baseline, PDE-CO reduces the error to $0.45\%$---a $17\times$ improvement.
This comparison uses the specified $N_d=11$ modal basis and regularization; it does not establish a universal error level for thin-plate modal sensing.
The difference in the plotted force patterns must therefore be interpreted together with the large PDE-CO force-vector error, rather than as evidence of uniquely identified loading.

\subsection{Bent pipe}\label{sec:7.2}

\begin{figure}[htbp]
\centering
\includegraphics[width=0.7\linewidth]{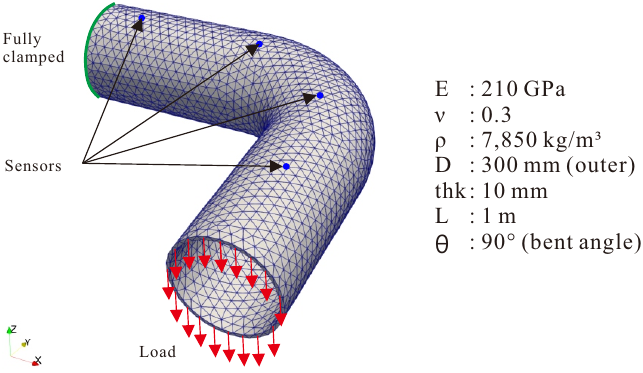}
\caption{Bent $90^\circ$ elbow pipe geometry, clamped end, eight acceleration sensors and loaded free-end cross-section. The numerical load is directed along the computational $y$ axis; the arrows mark the loaded end schematically without defining its harmonic phase. The annotated length $L=1$~m is the sum of the two 500~mm straight sections. The 300~mm-radius elbow contributes additional centreline arc length.}\label{fig:pipe_overview}
\end{figure}

The second example is a $90^\circ$ elbow pipe with outer radius $R_{\rm out} = 150$~mm, inner radius $R_{\rm in} = 140$~mm (wall thickness $t = 10$~mm), two straight segments of length $L = 500$~mm each, centerline bend radius $R_{\rm bend} = 300$~mm, and bend angle $90^\circ$; the total centerline path length is $2 \times 500 + 300\pi/2 \approx 1{,}471$~mm.
The material properties are $E = 210{,}000$~MPa, $\nu = 0.3$, and $\rho = 7.85 \times 10^{-9}$~tonne/mm\textsuperscript{3} (mm-tonne-s unit system).
The first straight segment is fully clamped; uniform positive $y$-direction nodal-force amplitudes are applied across the free-end cross-section of the second segment in the computational coordinates (see Figure~\ref{fig:pipe_overview}). This retains the boundary-face loading paradigm of the plate in a curved 3D structure, with a uniform rather than Gaussian nodal amplitude distribution.
The force is parameterized by $\boldsymbol{\theta} \in \mathbb{R}^{11{,}160}$ covering all free DOFs ($\mathbf{P} = \mathbf{I}$).
Eight acceleration sensors (24 DOFs) are placed on the outer surface --- four on the upper straight segment and four symmetrically on the lower straight segment.
The FE mesh has 11{,}300 tetrahedral elements (3{,}782 nodes, $N_f = 11{,}160$ free DOFs).
The excitation frequency is $\omega = 416.2$~rad/s (${\approx}66.2$~Hz), and the Tikhonov regularization parameter is $\alpha = 10^{-6}$.

\begin{figure}[htbp]
\centering
\includegraphics[width=0.95\linewidth]{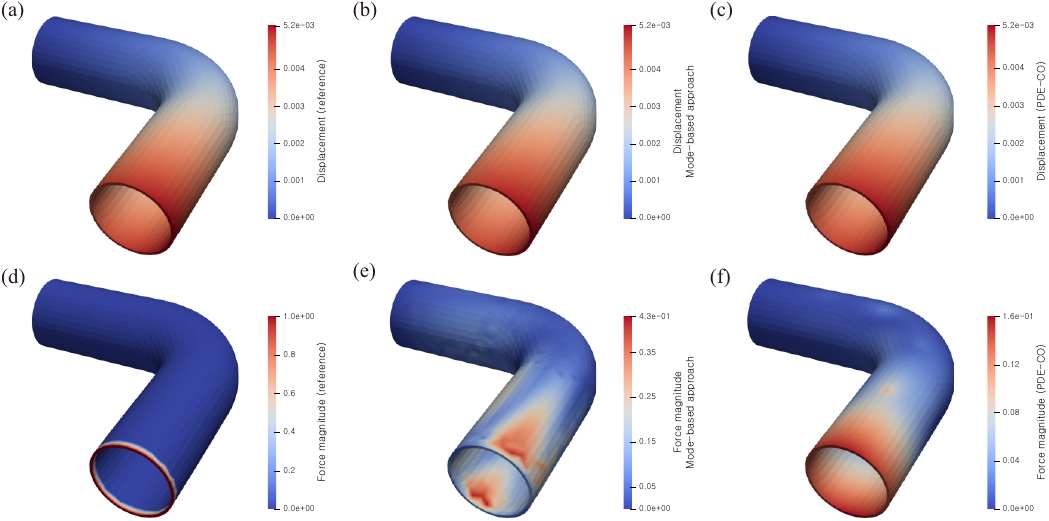}
\caption{Elbow pipe: original iterative reconstruction benchmark. Upper row: reference, mode-based approach and PDE-CO displacement magnitudes (mm). Lower row: corresponding nodal-force magnitudes (N). The inverse PDE-CO load space contains all 11{,}160 free DOFs. Each force panel retains its displayed colour scale, so the apparent load-region pattern is not a quantitative amplitude comparison. The PDE-CO nodal-force error is 92.56\%.}\label{fig:pipe_convergence}
\end{figure}

Figure~\ref{fig:pipe_convergence} presents the full-field comparison (six panels).
The upper row shows the displacement magnitude --- (a)~reference, (b)~the mode-based approach, (c)~PDE-CO.
Both methods capture the gross bending pattern (maximum at the free end, zero at the clamp), but PDE-CO reproduces the reference field more faithfully.
The lower row shows the recovered force magnitude distribution --- (d)~reference (distributed uniformly over the free-end cross-section), (e)~the mode-based approach, (f)~PDE-CO.
The force estimate from the mode-based approach (e) is spread along the pipe body with components outside the true free-end loading region.
The PDE-CO force estimate (f) shows a stronger visible concentration near the free-end cross-section, though its full nodal-force error is $\epsilon_f=92.56\%$. Distinguishing between concentrated and distributed loading from these sensor data remains ambiguous.

\begin{figure}[htbp]
\centering
\includegraphics[width=0.9\linewidth]{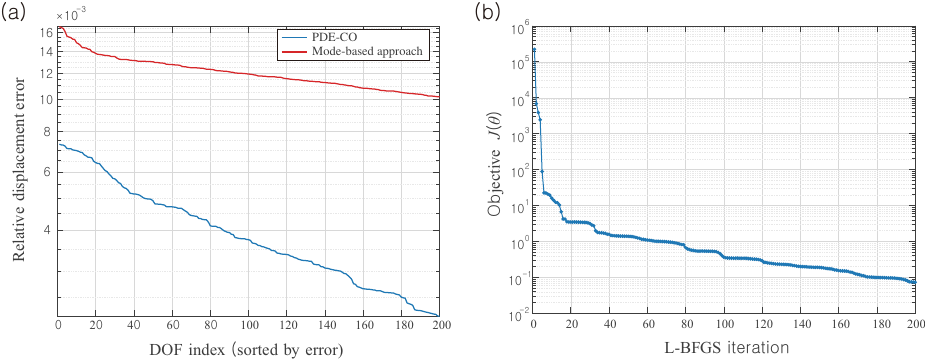}
\caption{Elbow pipe: original iterative benchmark diagnostics. (a) Largest 200 component errors $|\hat U_{f,i}^{\rm rec}-\hat U_{f,i}^{\rm true}|/\max_k|\hat U_{f,k}^{\rm true}|$, ranked independently for each method; the abscissa is error rank. Full-vector errors are 0.28\% for PDE-CO and 1.11\% for the mode-based approach. (b) Objective history through the imposed limit of 200 iterations. The optimiser did not report convergence.}\label{fig:pipe_disp_error}
\end{figure}

Figure~\ref{fig:pipe_disp_error} presents (a)~component displacement errors ranked independently for each method (top 200) and (b)~the L-BFGS objective history (200~iterations, ending at the imposed limit without reported convergence).
As reported in Table~\ref{tab:results_summary} (row 2), PDE-CO achieves $\epsilon_u = 0.28\%$ while the mode-based approach with 8 sensors yields $\epsilon_u = 1.11\%$.
Despite a substantial increase in problem size ($N_f = 11{,}160$ vs.\ $1{,}872$ for the plate), both methods achieve lower displacement errors in this example. The geometry, sensor layout, retained modes and load all differ, so this comparison alone does not isolate the cause.
The mode-based approach achieves $1.11\%$---a respectable result for a curved 3D structure.
PDE-CO reduces this to $0.28\%$, a $4.0\times$ reduction for the specified comparison, while the large force error remains unresolved.
The optimization wall time was 615~s for 200~iterations.

\subsection{Reactor pressure vessel (RPV)}\label{sec:7.3}

\begin{figure}[htbp]
\centering
\includegraphics[width=1\linewidth]{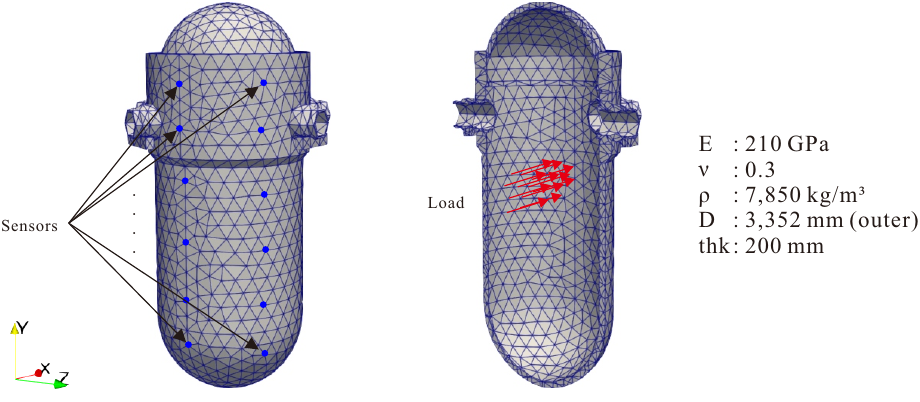}
\caption{Reactor pressure vessel schematic with 36 acceleration sensors on the outer surface and a localised load region on the inner surface. The numerical excitation is an $x$-directed Gaussian harmonic load amplitude. The annotated $D=3352$~mm is the upper cylindrical section's outer diameter, and $\mathrm{thk}=200$~mm is the original nominal thickness annotation. The CAD geometry has varying radii and wall thickness and defines the analysis geometry.}\label{fig:rpv_overview}
\end{figure}

The third example is a cylindrical reactor pressure vessel (RPV), the most complex case in both geometry and loading type. The CAD-based vessel is an illustrative numerical model; its geometry does not establish validation for an operating nuclear plant.
Unlike the plate and pipe, where the excitation is distributed on an accessible structural boundary (the free end), the RPV is excited by a localized Gaussian harmonic nodal load on the inner surface---a surface separated from the outer-surface sensors by the vessel wall. The excitation is a steady-state harmonic amplitude, not a simulated transient impact.
This shift from accessible end-face loading to inner-surface loading is physically motivated, as pressure-vessel loads can act on the inner wetted surface and may be difficult to measure directly. The prescribed $x$-directed nodal load is a benchmark loading, rather than a simulation of fluid pressure or a coupled fluid--structure process.
The identification of such loads from outer-surface sensor responses introduces a different load support and observation geometry from the preceding examples; its difficulty depends on the corresponding sensing operator.
The original geometry annotations give an inner cylindrical diameter $D_{\rm in}=3{,}150$~mm, an upper cylindrical outer diameter $D_{\rm out}=3{,}352$~mm and a nominal thickness $t=200$~mm (Figure~\ref{fig:rpv_overview}). These annotations refer to different parts or nominal descriptions; the CAD includes varying radii and wall thickness, and the three numbers do not define a single constant-thickness cylinder.
The model includes a cylindrical body with ellipsoidal/hemispherical heads, four nozzles (two set-in nozzles in the $\pm Z$ direction, two set-out nozzles at $\pm(X\!+\!Z)$, $60^\circ$ apart), and four external support pads.
The material properties are $E = 210{,}000$~MPa, $\nu = 0.3$, and $\rho = 7.85 \times 10^{-9}$~tonne/mm\textsuperscript{3} (steel, mm-tonne-s system).
The bottom faces of all four support pads are fully fixed (Dirichlet boundary condition).

Force parameterization is restricted to the inner surface free DOFs, $\boldsymbol{\theta} \in \mathbb{R}^{2{,}850}$ ($N_\theta = 2{,}850$, inner surface of the cylindrical body and hemispherical heads).
This encodes the assumed inner-surface load support; all three signed force components are allowed at the selected nodes, so the parameterization is more general than normal pressure loading.
The ground-truth force is a localized positive $x$-direction Gaussian harmonic nodal amplitude centered at mid-height on the $+X$ side of the inner cylindrical surface.

Thirty-six acceleration sensors ($6 \times 6$ grid, 6 height levels $\times$ 6 circumferential positions) are placed on the outer surface
(36 sensor nodes, 108 measured acceleration DOFs).
The FE mesh has 7{,}432 tetrahedral elements (2{,}446 nodes, $N_f = 7{,}206$ free DOFs; coarse mesh with $l_c = 150$--$400$~mm for tractable dense-solve times on CPU).
The excitation frequency is $\omega = 120.6$~rad/s ($\approx19.2$~Hz), and the Tikhonov regularization parameter is $\alpha = 10^{-12}$. The ratio of measured channels to force parameters is $108/2{,}850\approx3.8\%$; the load remains underdetermined despite the larger sensor count.

\begin{figure}[htbp]
\centering
\includegraphics[width=0.8\linewidth]{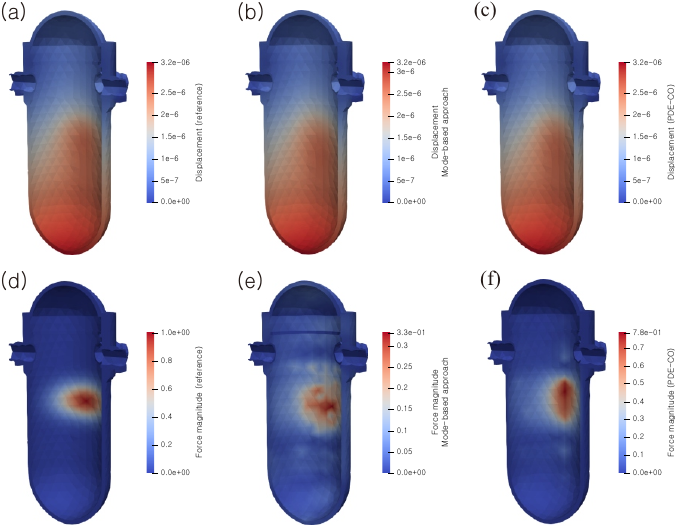}
\caption{Reactor pressure vessel: original iterative reconstruction benchmark. Upper row: reference, mode-based approach and PDE-CO displacement magnitudes (mm). Lower row: corresponding nodal-force magnitudes (N), shown on the inner surface for the load comparison. PDE-CO restricts the unknown load to 2{,}850 inner-surface DOFs. Force-panel scales differ; the PDE-CO full nodal-force error remains 48.02\% despite a compact visible load region.}\label{fig:rpv_convergence}
\end{figure}

Figure~\ref{fig:rpv_convergence} presents the full-field comparison for the RPV (six panels).
The upper row shows the displacement magnitude --- (a)~reference, (b)~the mode-based approach, (c)~PDE-CO.
All three show the same gross response pattern (large displacement at the lower hemispherical head, smaller at the nozzle ring), reflecting that both methods capture the dominant vibration shape.
The lower row shows the recovered force magnitude on the inner surface --- (d)~reference (compact Gaussian spot centered at mid-height on the cylindrical body), (e)~the mode-based approach, (f)~PDE-CO.
The mode-based approach (e) approximately identifies the loading region but produces a more diffuse distribution.
PDE-CO (f) produces a sharper visible spot near the reference load region; the individual colour scales and its full nodal-force error of $\epsilon_f=48.02\%$ preclude interpreting this as accurate recovery of the Gaussian profile.

\begin{figure}[htbp]
\centering
\includegraphics[width=0.9\linewidth]{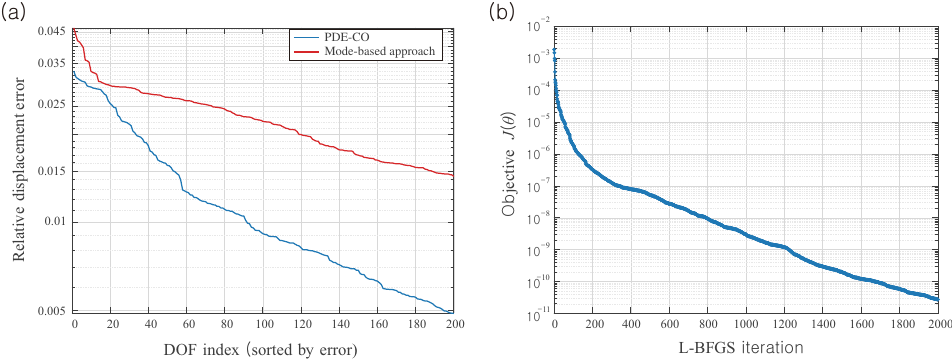}
\caption{Reactor pressure vessel: original iterative benchmark diagnostics. (a) Largest 200 component errors $|\hat U_{f,i}^{\rm rec}-\hat U_{f,i}^{\rm true}|/\max_k|\hat U_{f,k}^{\rm true}|$, ranked independently for each method; the abscissa is error rank. Full-vector errors are 0.885\% for PDE-CO and 1.86\% for the mode-based approach. (b) Objective history through 2{,}000 iterations, ending at the iteration limit without reported convergence. Continued decrease of the objective does not establish that additional iterations would improve the true force or unmeasured response.}\label{fig:rpv_disp_error}
\end{figure}

Figure~\ref{fig:rpv_disp_error} presents (a)~component displacement errors ranked independently for each method (top 200) and (b)~the L-BFGS objective history through 2{,}000~iterations.
The L-BFGS optimizer reached its imposed limit of 2{,}000~iterations without reporting convergence. The displacement error at termination was $\epsilon_u=0.885\%$; continued decrease of the objective does not establish that further iterations would improve the true force or unmeasured response.
As reported in Table~\ref{tab:results_summary} (row 3), PDE-CO achieves $\epsilon_u = 0.885\%$ while the mode-based approach with 36 sensors yields $\epsilon_u = 1.86\%$.
The mode-based approach achieves $1.86\%$ for this vessel geometry with 36 triaxial sensors. This is a response estimate; the channel-to-force-parameter ratio does not establish unique load recovery.
PDE-CO reduces this to $0.885\%$ ($2.1\times$ improvement), and does so despite the forcing being on the inner surface, not directly measured by the outer-surface sensors, and the optimizer not yet having converged.
The apparent difference in load localization is accompanied by a substantial PDE-CO force-vector error; the direct-solution and reliability studies below examine this distinction quantitatively.
The optimization required 15{,}294~s for 2{,}000 iterations (7.6~s per iteration), reflecting the dense linear solve cost at this problem scale.

The original force and displacement arrays, optimization summaries and plotting inputs underlying Figures~\ref{fig:plate_convergence}--\ref{fig:rpv_disp_error} are included in Supplementary Material S1. The following analyses retain these benchmarks and add independently computed direct solutions, output diagnostics and perturbed conditions.

\subsection{Noise sensitivity and output reliability}\label{sec:7.validation}
Noise sensitivity is evaluated for all three structures using matched models at $\omega=0.5\omega_1$ with mass-proportional damping $\mathbf C_{d,ff}=0.02\omega_1\mathbf M_{ff}$. The complex sensor-acceleration vector is perturbed at 0, 1, 3 and 5\% of the noise-free vector's Euclidean norm, with 30 realizations per nonzero level shared by the two estimators.

The mode-based approach uses the same retained number of modes as in the original examples. PDE-CO uses 78 tip-face force components for the plate, all 11{,}160 free force components for the pipe and 2{,}850 inner-surface components for the RPV. Both methods select regularization by the discrepancy rule using the prescribed noise norm, with penalties on forces and modal coordinates, respectively. The PDE-CO values are direct minimizers of the regularized linear problem; the separate plate AD check agrees to $2.10\times10^{-9}$ in relative force norm.

\begin{figure}[htbp]
\centering
\includegraphics[width=\linewidth]{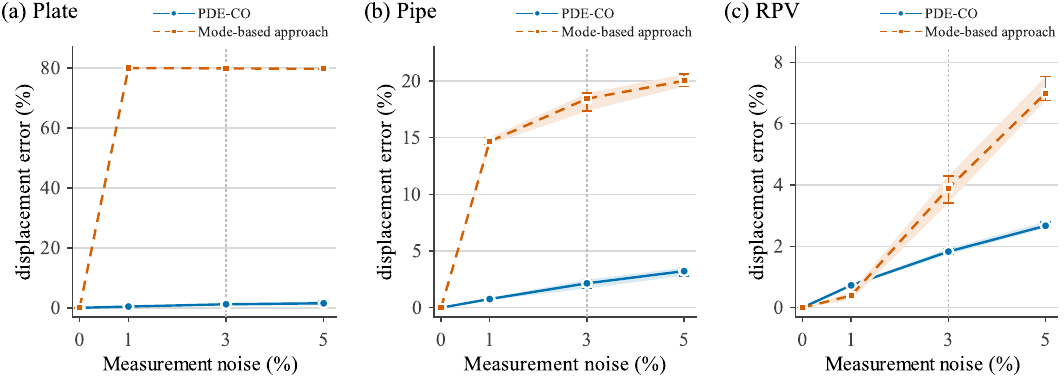}
\caption{Effect of measurement noise on displacement reconstruction for (a) plate, (b) pipe and (c) RPV. The ordinate shows the increase in relative displacement error at non-sensor nodes from each method's zero-noise baseline, $\Delta\epsilon_{u,\mathrm{ns}}(\eta)=\epsilon_{u,\mathrm{ns}}(\eta)-\epsilon_{u,\mathrm{ns}}(0)$, expressed in percent; $\eta$ denotes the measurement noise level. Markers show medians; bands and whiskers show interquartile ranges over 30 realizations at each nonzero noise level. Vertical axes are linear and use different ranges. Dotted lines mark the 3\% condition compared in Table~\ref{tab:noise_displacement}.}\label{fig:validation_summary}
\end{figure}

Table~\ref{tab:noise_displacement} reports the relative displacement error at non-sensor nodes, $\epsilon_{u,\mathrm{ns}}$. It uses the same norm ratio as $\epsilon_u$, with both displacement vectors restricted to free DOFs at nodes without sensors; Sections~\ref{sec:7.1}--\ref{sec:7.3} report the full-vector error. The difference $\Delta\epsilon_{u,\mathrm{ns}}=\epsilon_{u,\mathrm{ns}}(3\%)-\epsilon_{u,\mathrm{ns}}(0\%)$ is expressed in percent. At 3\% noise, $\Delta\epsilon_{u,\mathrm{ns}}$ is 1.18\%, 2.15\% and 1.83\% for PDE-CO, compared with 79.86\%, 18.44\% and 3.90\% for the mode-based approach. The smaller increases in all three cases support greater displacement robustness for PDE-CO under the tested priors and regularization choices. Figure~\ref{fig:validation_summary} shows the full noise sweep; this ordering is not uniform at every noise level, as the mode-based approach gives a smaller increase for the RPV at 1\% noise.

\begin{table}[htbp]\centering\small
\caption{Relative displacement error at non-sensor nodes before and after adding 3\% acceleration noise. Errors and differences are in \%; $\Delta\epsilon_{u,\mathrm{ns}}$ is the difference from the zero-noise error, computed before rounding. Nonzero-noise values are medians over 30 realizations. The matched-model frequency, damping and regularization-selection rule are held consistent across noise levels.}\label{tab:noise_displacement}
\begin{tabular}{llrrr}\toprule
Case & Method & 0\% noise & 3\% noise & $\Delta\epsilon_{u,\mathrm{ns}}$ (\%)\\\midrule
Plate & PDE-CO & 0.001 & 1.18 & 1.18 \\
 & Mode-based approach & 7.73 & 87.59 & 79.86 \\
\addlinespace
Pipe & PDE-CO & 0.24 & 2.39 & 2.15 \\
 & Mode-based approach & 1.11 & 19.55 & 18.44 \\
\addlinespace
RPV & PDE-CO & 0.89 & 2.72 & 1.83 \\
 & Mode-based approach & 1.87 & 5.77 & 3.90 \\
\bottomrule\end{tabular}
\end{table}

At 3\% noise, an input-based ROM with the same force prior gives displacement errors of $1.17\%$, $2.39\%$ and $2.58\%$, comparable to PDE-CO. The comparison therefore concerns the force prior and estimation procedure as well as the forward representation.

The reference loads are confined to boundary regions: Gaussian nodal loads on the plate tip and RPV inner surface, and uniform nodal loading across the pipe free-end cross-section. Reconstructing these loads as more broadly distributed nodal forces can contribute to large relative nodal-force errors even when the displacement error remains small. At 3\% noise, the PDE-CO force errors remain $67.19\%$, $95.02\%$ and $66.83\%$ for the plate, pipe and RPV. A separate plate ambiguity check also gives effectively identical sensor readings for loads differing by $100\%$, with displacement and cell-average strain differences of $0.840\%$ and $31.30\%$. These observations lead to the conclusion that robustness in displacement reconstruction does not establish unique or accurate load recovery.

\section{Conclusions}\label{sec:8}
Classical mode-based virtual sensing primarily focuses on displacement reconstruction, but load data are also essential for fatigue and fracture assessment, damage detection and remaining-life prediction in SHM and PHM.
To address this need, the proposed PDE-CO formulation jointly reconstructs harmonic loads and the full displacement field from sparse acceleration measurements while enforcing finite element equilibrium.
The framework separates offline assembly of the structural operators from online reconstruction, where L-BFGS-B optimization solves a regularized least-squares problem using gradients computed by automatic differentiation.
Forward and gradient evaluations can be accelerated on a GPU to support reconstruction with a large number of load parameters.
Across the plate, pipe and reactor pressure vessel benchmarks, PDE-CO achieved substantially lower displacement errors than the mode-based approach with the original retained number of modes.
When 3\% acceleration noise was added, PDE-CO also exhibited smaller increases in displacement error at non-sensor nodes in all three structures, indicating greater noise robustness under the tested matched-model conditions.
A limitation of the current formulation is that concentrated loads are estimated as distributed loads.
This framework provides a basis for incorporating joint load and displacement reconstruction into SHM and PHM systems.
The present study focuses on developing the theoretical and computational framework using synthetic numerical data; future experimental validation is needed to assess its performance under practical monitoring conditions.

\section*{Author Contributions}
\textbf{Minjae Kim}: Data curation, Visualization.
\textbf{Jaehwan Jeong}: Data curation, Visualization.
\textbf{Seulki Lee}: Writing---review \& editing.
\textbf{Jaemin Kim}: Conceptualization, Methodology, Software, Validation, Formal analysis, Investigation, Resources, Data curation, Writing---original draft, Writing---review \& editing, Visualization, Supervision, Project administration, Funding acquisition.

\bibliographystyle{elsarticle-num}
\bibliography{references}

\end{document}